\documentclass[aip]{revtex4}
\usepackage{graphicx}
\usepackage{amsmath}

\begin{document}

\title[Search for an additional neutrino mass eigenstate  from ''Troitsk nu-mass'' ]{ A search for an additional neutrino mass eigenstate  in 2 to 100 eV region from ''Troitsk nu-mass'' data -- detailed  analysis}

\author{ A.\,I.\,Belesev}
\author{ A.\,I.\,Berlev}
\author{ E.\,V.\,Geraskin}
\author{ A.\,A.\,Golubev}
\author{ N.\,A.\,Likhovid}
\affiliation{Institute for Nuclear Research of Russian Academy of 
Sciences, 117312 Moscow, Russia}
\author{ A.\,A.\,Nozik}
\email{nozik@inr.ru}
\affiliation{Institute for Nuclear Research of Russian Academy of 
Sciences, 117312 Moscow, Russia}
\affiliation{Moscow Institute of Physics and Technology, 141700 
Dolgoprudny, Russia}
\author{ V.\,S.\,Pantuev}
\email{pantuev@inr.ru}
\author{ V.\,I.\,Parfenov}
\author{A.\,K.\,Skasyrskaya}
\affiliation{Institute for Nuclear Research of Russian Academy of 
Sciences, 117312 Moscow, Russia}

\begin{abstract}
In this paper we present the details of our previously published results for  a search for an additional neutrino mass state in $\beta$-electron spectrum from the Troitsk nu-mass experiment. Here we present steps of the analysis, show a set of likelihood functions obtained for each additional heavy mass value. We demonstrate how systematic errors were estimated. We also compare our results with those published recently for a similar analysis for Mainz data and try to explain why there is a factor of 2-3 difference in the sensitivity for an additional heavy mass.
\end{abstract}


\maketitle

\section{Introduction}
Recently we have published a short paper on upper limits on an additional neutrino mass eigenstate in 2 to 100 eV region~\cite{short} obtained after reanalysis of our data on the direct electron antineutrino mass measurements in Tritium $\beta$-decay in the Troitsk nu-mass experiment~\cite{our_nu_e}.  The group led by V. M. Lobashev took these data in the period of 1997-2004. The same file set and the same analysis framework as for the electron anti neutrino mass were used. A search was performed for any sign of an additional neutrino mass state in the $\beta$-electron spectrum. Such a state with a finite mass would exhibit itself as a kink in the spectrum. The analysis drew much attention to the fact how data were analyzed, how systematic errors were estimated and why our limits appeared to be by a factor of 2-3 better than those from the similar analysis for Mainz data~\cite{mainz-sterile}. Here we present all details of the analysis and make a direct comparison with Mainz results.
\section{Raw spectrum analysis}
All details on the experimental setup, data taking, analysis and corrections  are published in Ref.~\cite{our_nu_e}. We use exactly the same data set with all the same data corrections. The only difference is that a likelihood function was used here for parameter estimations. To keep description on a smooth track, we repeat an introductory part from~\cite{short}. 

In accordance with Ref.~\cite{our_nu_e}, the spectrum of electrons in tritium  $\beta$-decay is the following:

\begin{equation}
S(E,E_0,m_\nu^2) = NF(E)(E+m_{e})p_e(E_0-E)\cdot
\sqrt{(E_0-E)^2-m_\nu^2}
\label{eq:one}
\end{equation}

where $N$ is the normalization constant, $F(E)$ - the so called Fermi-function responsible for electrostatic interaction between electron and nucleus,  $E$ and  $p_e$ stand for the electron energy and momentum, $E_0$ - for the beta-spectrum endpoint and $m_\nu$ - for the neutrino mass.
After decay of a tritium nucleus the primary molecule of T$_2$ becomes a molecule of T$^3$He. Often, with a probability of about 43\%,  T$^3$He$^+$ does not go to its ground state, thus we have to sum over all molecule final states $i$ and eq.~\ref{eq:one} 
should be replaced by the sum:

\begin{equation}
T(E,E_0,m_\nu^2)=\sum_i S(E, E_0-\varepsilon_i,m_\nu^2)\cdot P_i , 
\label{eq:states}
\end{equation}

where $\varepsilon_i$ is the energy of the excited state and $P_i$ is its probability, the sum of  $P_i$ equals one.  
Finally, we get the following expression for the experimental integrated electron spectrum versus retarding potential on the spectrometer electrode $V$:

\begin{equation}
Sp(V)=N\cdot \int \big[T(E,E_0,m_{\nu}^2)\otimes Y(E) \big]\cdot R(V,E)dE + bkgr ,
\label{eq:spectrum}
\end{equation}

where $T(E,E_0,m_{\nu}^2)$ is the electron spectrum from the $\beta$-decay, Eq.~\ref{eq:states}; $Y(E)$ is the energy loss spectrum and $R(V,E)$ is the resolution function (see~\cite{our_nu_e} for details),  and $bkgr$ is the experimental background. 

In case of neutrino mixing, for the effective electron neutrino one can write $\mid\nu_e\rangle=\sum\limits_i U_{ei}\mid\nu_i\rangle$, where $U_{ei}$  are the mixing matrix elements. We restrict ourselves to one additional heavy neutrino ($i=4$). From neutrino oscillation results it is known that the mass splitting between active neutrinos is much less than one electron-volt. Thus, masses of "normal" eigenstates are probably negligibly small,  and one can assume $m_1=m_2=m_3=0$. Consequently, the electron spectrum with one additional heavy neutrino component can be written as follows:

\begin{equation}
{S(E) = NF(E)(E+m_{e})p_e(E_0-E) 
\cdot \biggl[ U^2_{e4}\sqrt{(E_0-E)^2-m_4^2}+(1-U^2_{e4})(E_0-E)\biggr],}
\label{eq:funct}
\end{equation}

where $U^2_{e4}$ is the fraction of the heavy neutrino in the electron neutrino and $m_4$ is the mass of the heavy neutrino eigenstate (here by $U^2_{e4}$ we denote $ \mid U_{e4} \mid ^2$). In other words, we fit the spectrum with an assumption that its  major component has a relative amplitude $1-U^2_{e4}$ and is attributed to zero neutrino mass, besides there is an additional feature with the relative amplitude $U^2_{e4}$ for heavy mass $m_4$. One has to stress that~Eq.\ref{eq:funct} represents a different functional compared to the standard one without an additional term for $U^2_{e4}$ and the role of systematic errors could be different. It is also worth mentioning that in a usual notation for the neutrino oscillations parameter $sin^2(2\theta)$~\cite{oscill}, at small $U^2_{e4}$ there is an approximate relation $sin^2(2\theta) \approx 4 U^2$.
\section{Likelihood function}
\subsection{Construction of likelihood function}
\begin{figure}[h]
\parbox{0.48\linewidth}{\includegraphics[width = \linewidth]{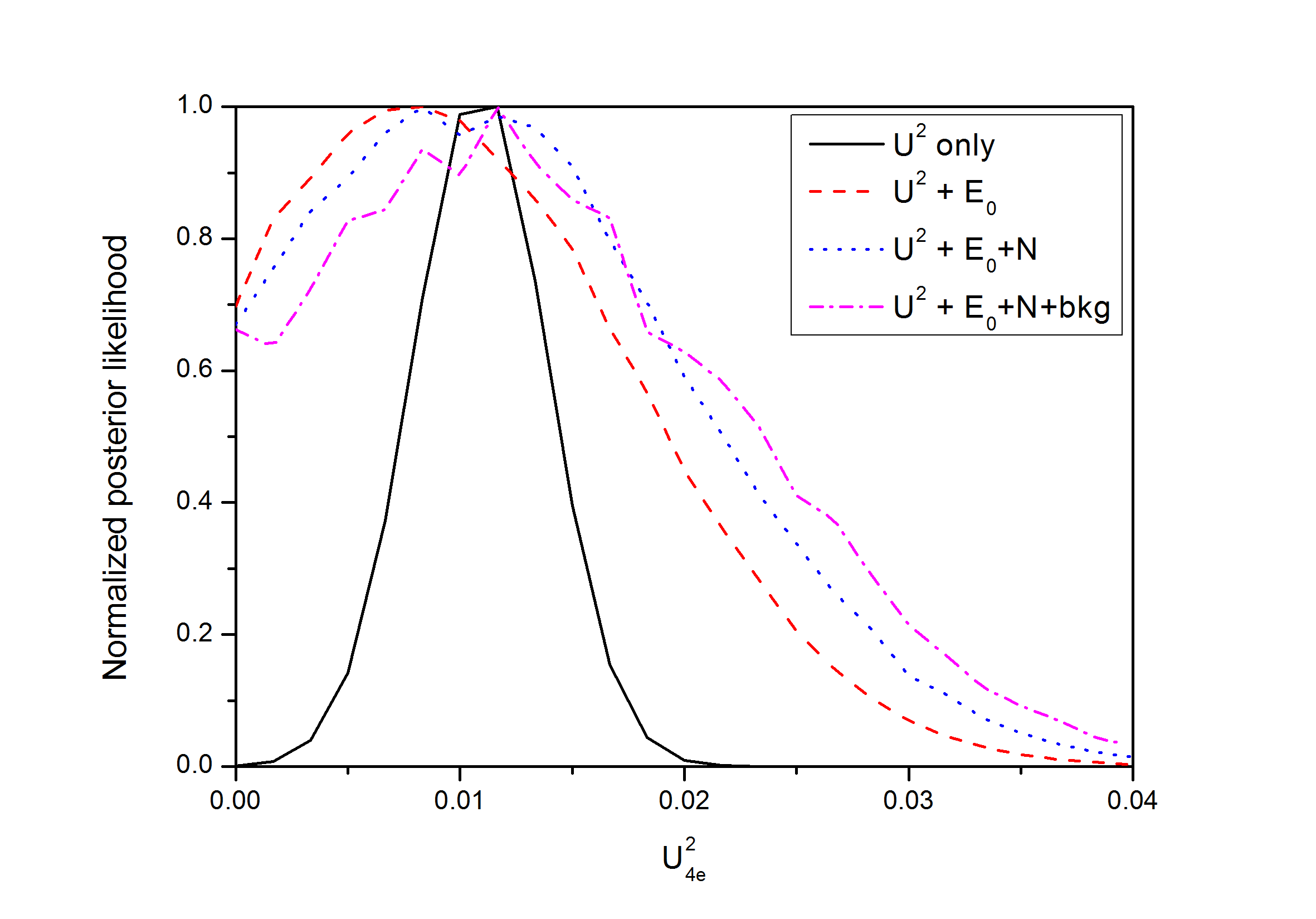}}\parbox{0.48\linewidth}{\includegraphics[width = \linewidth]{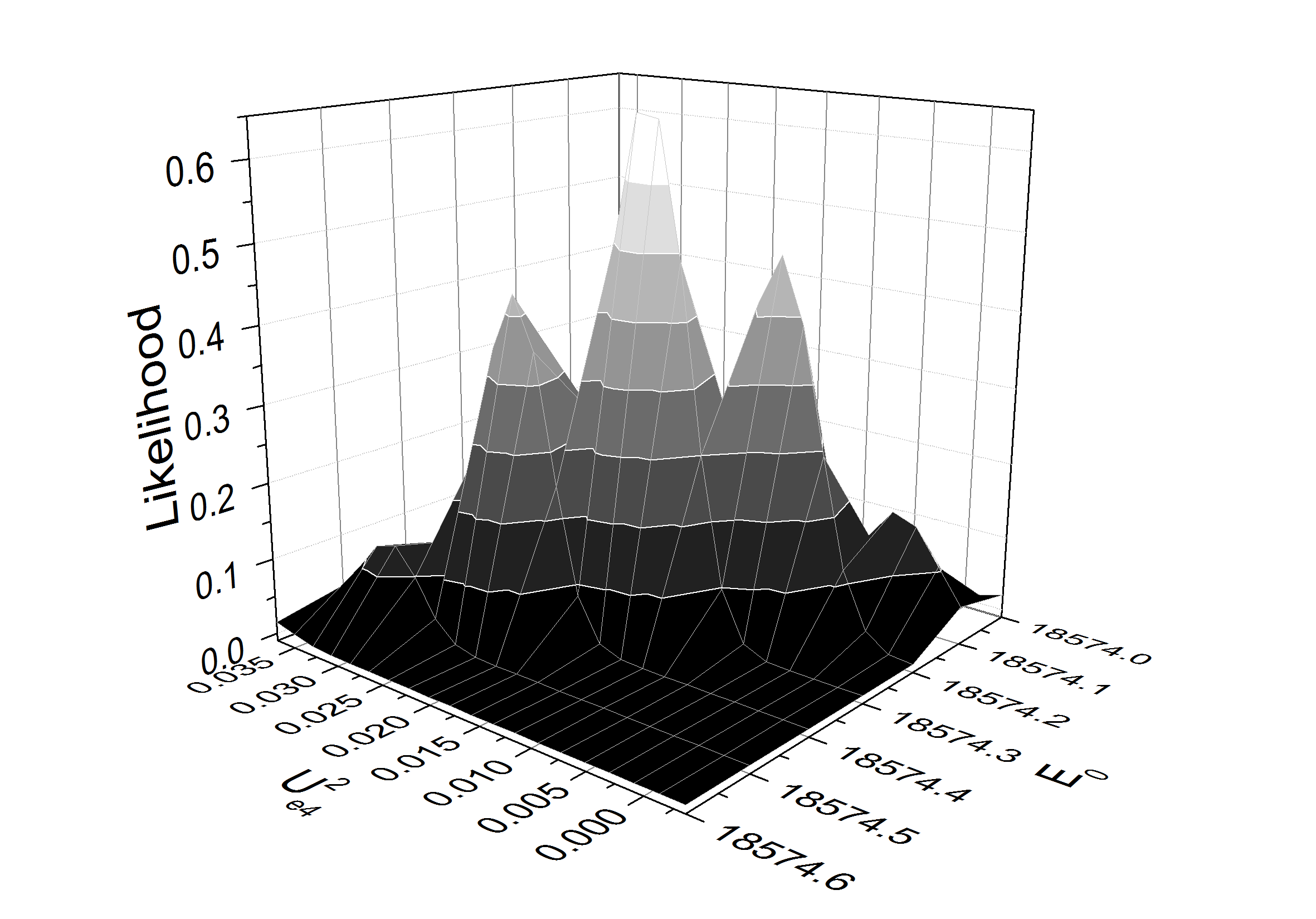}}\\ 
\parbox{0.48\linewidth}{\caption{Likelihood functions integrated sequentially over different parameters for $m_4$=50~eV.}\label{fig:step_by_step_integration}}
\hfill
\parbox{0.48\linewidth}{\caption{Two-dimensional plot of likelihood function for one of the runs for parameters $U^2_{e4}$ and $E_0$ at $m_4$=50 eV.}\label{fig:like2d}}
\end{figure}

To get an upper limit for $U^2_{e4}$ the Bayesian approach has been used for the parameter estimation.  For each run the probability calculation procedure is the following: 
\begin{enumerate}
\item At first, we set $U^2_{e4}$  to zero and fit three spectrum parameters: $E_0$, $N$ and $bkg$. This is required to get a precise region for the additional parameters. One must note that, while $E_0$ is a physical value and should not change from run to run, in practice it depends on the spectrometer calibration, can vary for different data sets and is used as a free parameter.
\item Next step we set $m_4^2$ and construct a four-dimensional likelihood function $L(U^2_{e4}, E_0, N, bkg)$. The likelihood, by definition, is a probability to describe the given data with the given spectrum parameters.  Since  in our case the number of events at each data point is always larger than 100, we  substitute the Poisson distribution with the Gaussian function.  So, if $\mu(V,U_{e4}^2,E_0,N,bkg)$ is an estimated total count at the spectrometer voltage $V$, then the resulting likelihood function over all points in the spectrum for one run can be written in the following way: 
\begin{equation}
L(U_{e4}^2,E_0,N,bkg) = \prod\limits_i \frac{1}{\sqrt{2 \pi \sigma_i^2}} e^{-\frac{(X_i - \mu_i)^2}{2 \sigma_i^2}},
\label{eq:likelihood}
 \end{equation}
where $\mu_i = \mu(V_i,U_{e4}^2,E_0,N,bkg)$ is the estimate of the spectrum value at the point $i$ at the spectrometer voltage $V_i$ and $X_i$ is the experimental count number at this point.
\item We are not interested in parameters $ E_0$, $N$ or $bkg$. but we can't just take projection of $L(U^2_{e4}, E_0, N, bkg)$ on $U^2_{e4}$  or make a cut because the all parameters are correlated at different strengths. In order to take into account correlations, integration is made by the Monte-Carlo procedure. We marginalize the likelihood function over all non-essential parameters by integrating  them over 3-dimensions: $L(U^2_{e4}) = \int\limits_{E_0}\int\limits_{N}\int\limits_{bkg} L(U^2_{e4}, E_0, N, bkg)$. The margins for integration are taken between plus and minus four standard deviations of each parameter around the most probable values obtained in step 1 . By this integration covers the most part of the parameter ranges of $L(U^2_{e4}, E_0, N, bkg)$ which have been constructed at step 2. In Fig.~\ref{fig:step_by_step_integration} we illustrate how the likelihood function gets wider would integration be done over variables step by step. Calculation of the likelihood function for one set of parameters is greatly computation time consuming. For each value  of  $m_4^2$ we dice 70 000 times. In some cases to check stability of the result this number was changed to 30 000 and 150 000. The values of $L(U^2_{e4})$ are saved in the table with corresponding values of $U^2_{e4}$. The values of $U^2_{e4}$, for which $L(U^2_{e4})$ were calculated, were defined in the following way. At first there were preliminary estimations of the maximum value of $L(U^2_{e4})$ and a range where it had a meaningful value. Then 25 points of $U^2_{e4}$ at the same distance were selected in the interval from zero to the point where $L(U^2_{e4})$ goes down to at least $10^{-3}$ of its maximum value. For example, such intervals were 0 to 1, 0 to 0.5, 0 to 0.1 for masses $m_4$=3 eV, 5 eV, 10 eV, respectively.
\item We repeat the procedure from step 2 for different values of $m_4^2$.
\end{enumerate}

We have to note that while the marginal likelihood function obtained by integration over three variables has a quasi-Gaussian shape, the sub-dimensional likelihood function which is obtained by integration over one or two parameters may not be Gaussian. An example could be seen in Fig. \ref{fig:like2d}, where integration is done over normalization and background parameters, $N$ and $bkg$ . This is an untypical case, but it demonstrates a potential problem if you try to fit, for example, the function projection. The maximum likelihood method automatically solves such a problem.    

During all the calculations we presume that $0 \leq U^2_{e4} \leq 1$. The final posterior probability $L$ for parameter $U^2_{e4}$ is calculated as a product of posterior probabilities $L_k$ for different experimental runs ($L(U^2_{e4}) = \prod L_k(U^2_{e4})$). At last, an upper limit for each value of $m_4^2$ has been found by solving the equation:

\begin{equation}
\frac{\int\limits_0^{limit}L(U^2_{e4})}{\int\limits_0^{1}L(U^2_{e4})} = \alpha,
\label{eq:limit}
\end{equation}

where $\alpha$ is the required confidence level, namely 0.95, and the $limit$ is the relevant value of  $U^2_{e4}$ .

\begin{figure}[h]
\center\includegraphics[width = 80 mm]{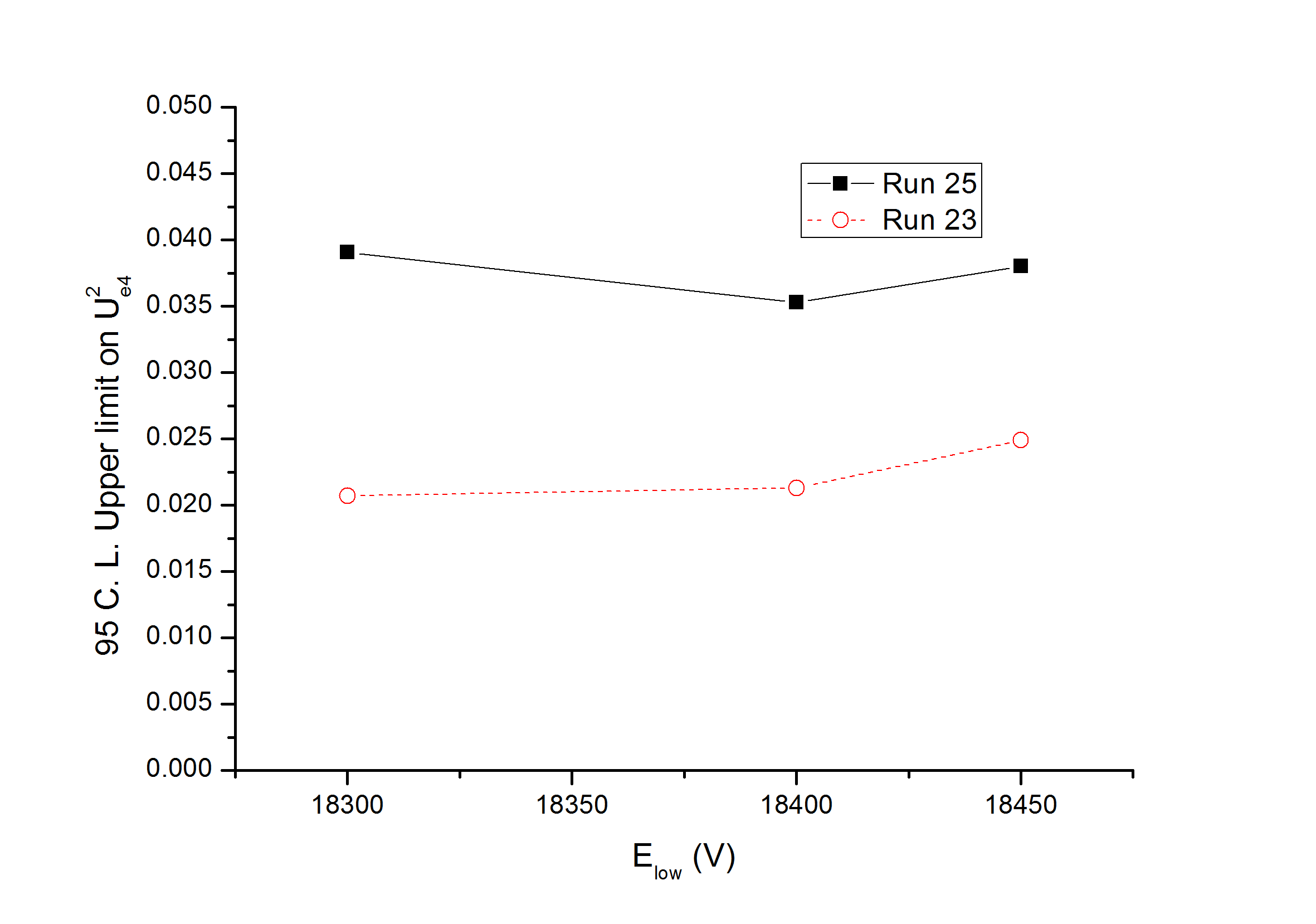}
\caption{The upper limit dependence on the choise of $E_{low}$ for two runs with different statistics; $m_4 = 50\,eV$.}
\label{fig:elowtest}
\end{figure}

In the current analysis we used the data in which the spectrometer electrode potential was higher than $E_{low} = 18400~V$, which corresponds to about a $175~V$ range. We applied such restriction because for lower spectrometer voltages there are additional systematic errors arising from the detector dead time and registration efficiency. We also checked that the extension of the data range to $E_{low} = 18300~V$ or $18450~V $ does not dramatically change the result, Fig.~\ref{fig:elowtest}.

\subsection{Calculation details}

Calculation of a likelihood function for one set of parameters takes approximately 0.5 second on a modern computer. While being fast enough for different fitting methods this time is too long to implement the Monte-Carlo integration procedure described earlier. If one calculation takes 0.5 second, then 70000 calculations require 35000 seconds (almost ten hours). The most time consuming part is the calculation of integral in Eq.~\ref{eq:spectrum}  with substitution of $S(E)$ from Eq.~\ref{eq:funct} (we will designate this integral as $I$). One can note that $I$ does not depend on $N$ and $bkg$. For small $\Delta E$ (the difference between the upper and lower boundaries $\pm4 \sigma$ for parameter $E_0$  is about 0.4 eV for each single run) the dependence on the endpoint energy $E_0$ can be  described as $I(E, E_0) = I(E+ \Delta E, E_0+\Delta E)$. Taking this into account we can simplify the calculation procedure. For each $U^2$ we calculate $I$ for the given value of $U^2$ and $\hat{E_0}$ which is $E_0$ obtained from the fit with $U^2  = 0$. We interpolate $I$ with a linear approximation with 300 nodes and obtain piecewise-defined function $\hat{I}(E, \hat{E_0})$. Each time we need to calculate the value $I(E, E_0)$, we'll just use the value $\hat{I}(E - E_0 + \hat{E_0}, \hat{E_0})$ instead.

\begin{figure}[htb]
\begin{minipage}[t]{0.48\linewidth}
\includegraphics[width=1.\textwidth]{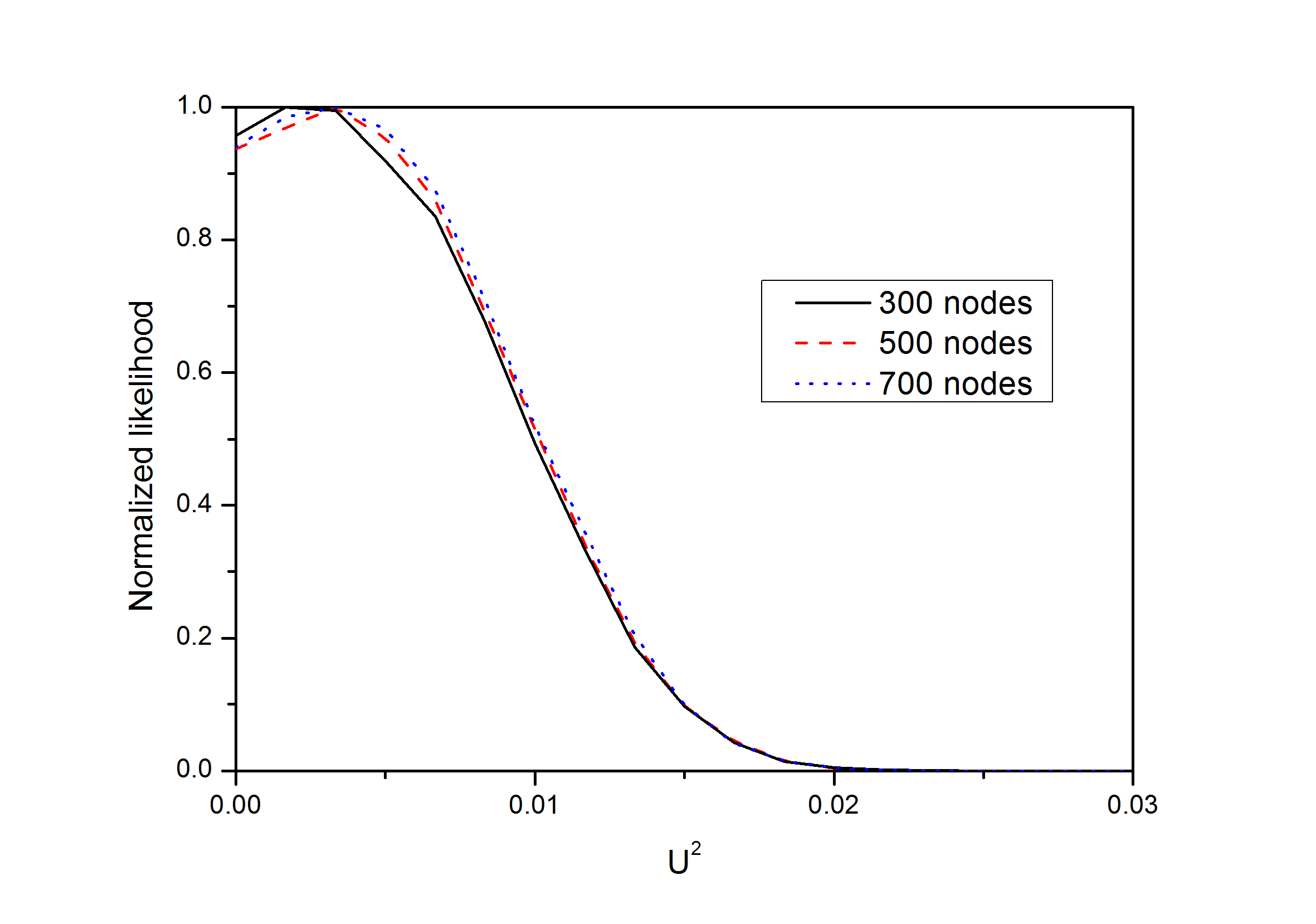}
\end{minipage}
\hfill
\begin{minipage}[t]{0.48\linewidth}
\includegraphics[width=1.\textwidth]{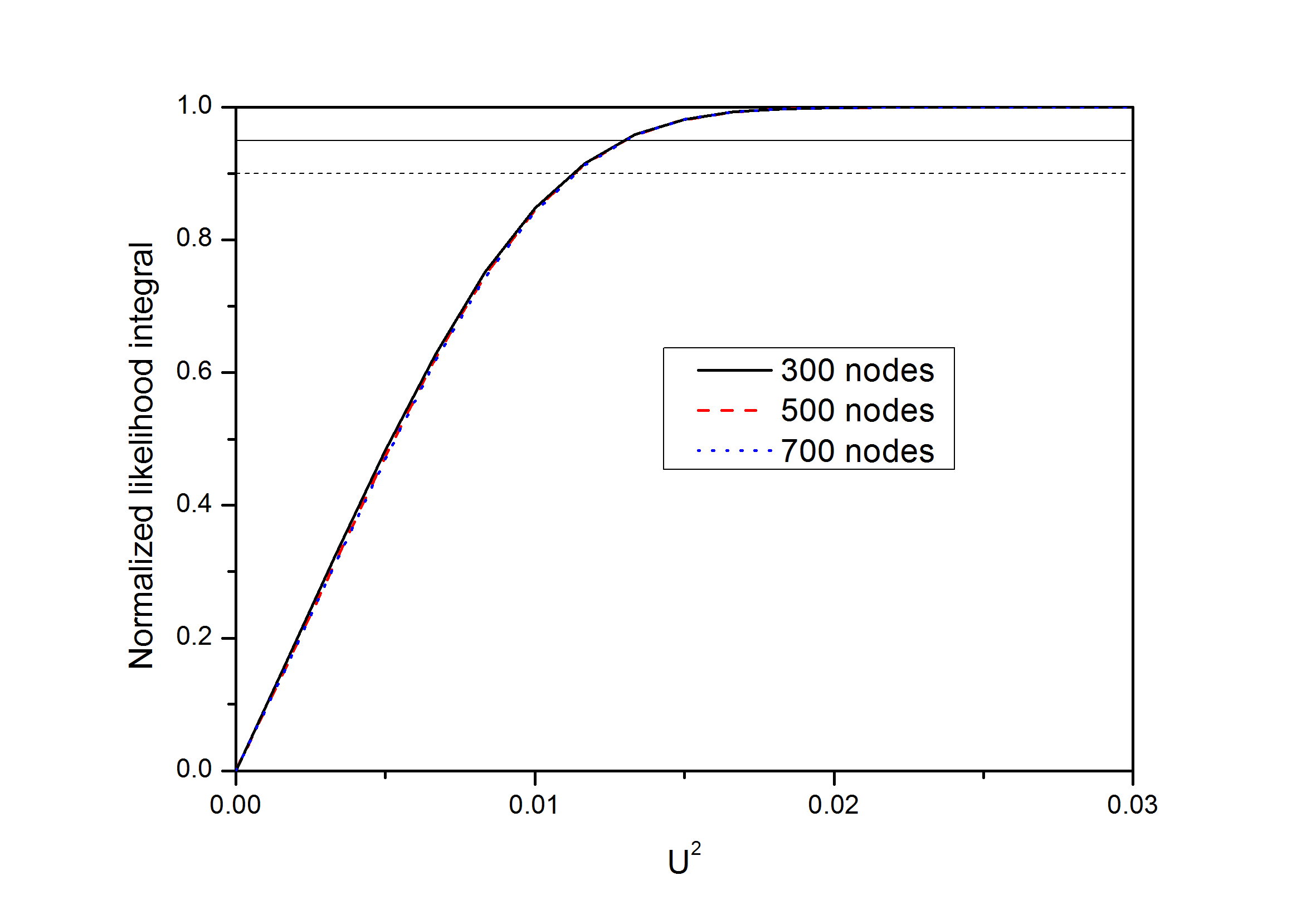}
\end{minipage}
{\caption{On the left: comparison of the likelihood functions obtained with a different number of interpolation nodes. On the right: integral of these likelihood functions.}
\label{fig:splinetest}}
\end{figure}

We've checked that the interpolation procedure does not affect the result. First, we proved that by increasing the number of nodes don't change the result. The comparison for one of the experimental runs is shown in Fig.~\ref{fig:splinetest}. We also checked that interpolation does not change the likelihood shape and the resulting curve coincides with the one obtained by direct calculation.

\subsection{Systematic errors estimation}

It is worth stressing that all statistical errors and correlations are already incorporated during construction and calculation of likelihood functions, Eq.~\ref{eq:likelihood}, and included in the upper limit estimation by Eq.~\ref{eq:limit}. During the experiment we don't know exactly or cannot control some of the experimental conditions. Such uncertainties define our systematic errors and should be carefully evaluated.
The correct way to treat a systematic parameter in Bayesian approach is to marginalize the likelihood function over systematic parameter  with a prior probability corresponding to the expected spreading of systematic parameter. According to the standard procedure for Bayesian treatment of nuisance parameters the marginal likelihood was calculated as:
\begin{equation}
L(U^2)=\int L(U_{e4}^2,X)\pi(X) dX,
\label{eq:system}
\end{equation}

A flat distribution $\pi(X)$ was assumed for a systematic parameter within its maximum range. We also provided a conservative estimation of systematic influence which was made in the following way:
\begin{enumerate}
\item For each value of $m_4^2$ we get the combined likelihood function for all experimental runs  with most probable apparatus parameters  and calculate an upper limit on the mixing parameter.
\item Then we set an apparatus parameter (e. g. source thickness) to its systematic boundary (upper or lower depending on which one provides a higher upper limit ) and repeat the analysis procedure for the shifted value.
\item The maximum deviation of an upper limit from the value obtained in step 1 shows the maximum effect from apparatus parameters.  
\end{enumerate}

\begin{figure}[h]
\begin{minipage}[t]{0.48\linewidth}
\includegraphics[width=1.\textwidth]{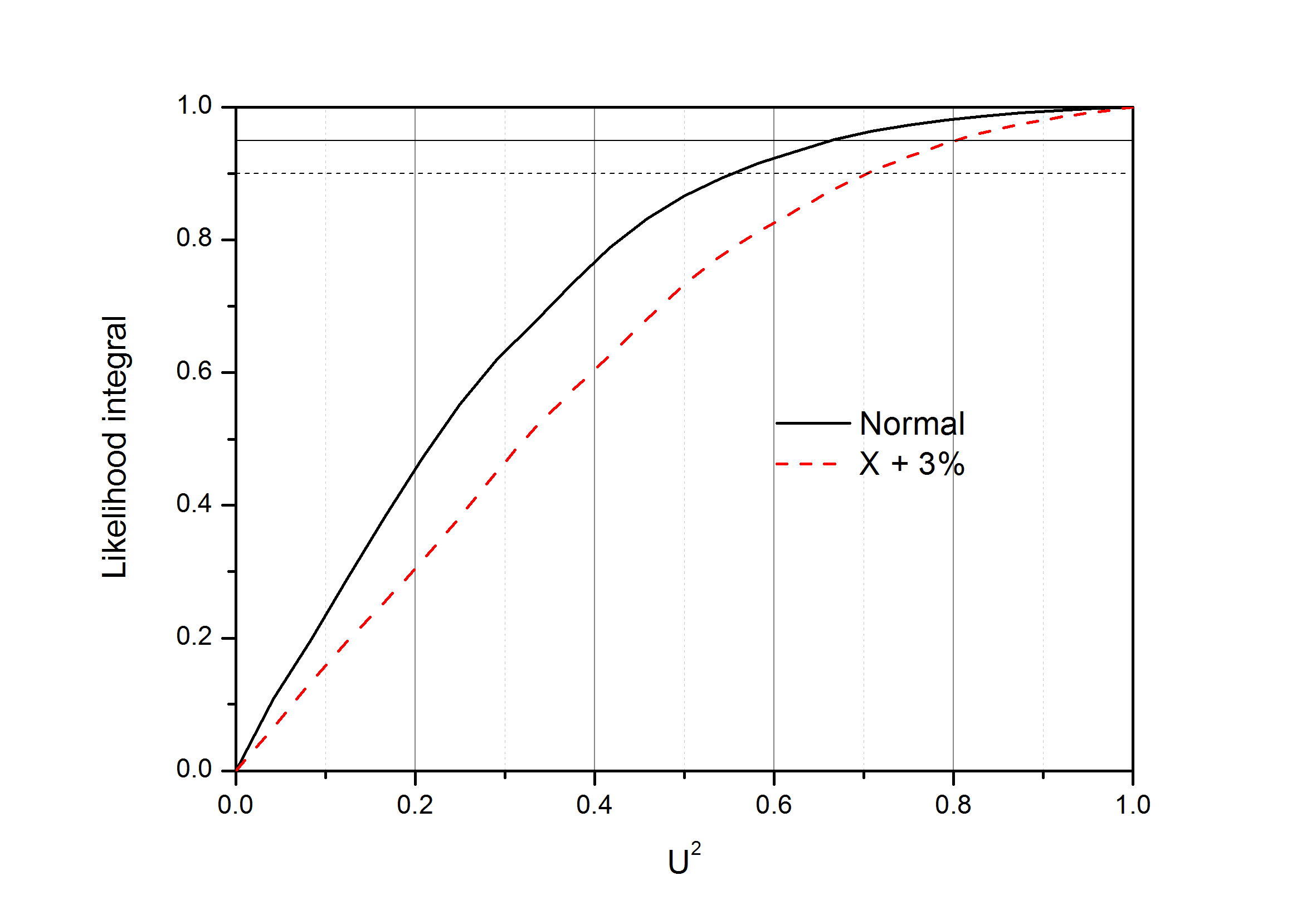}
\end{minipage}
\hfill
\begin{minipage}[t]{0.48\linewidth}
\includegraphics[width=1.\textwidth]{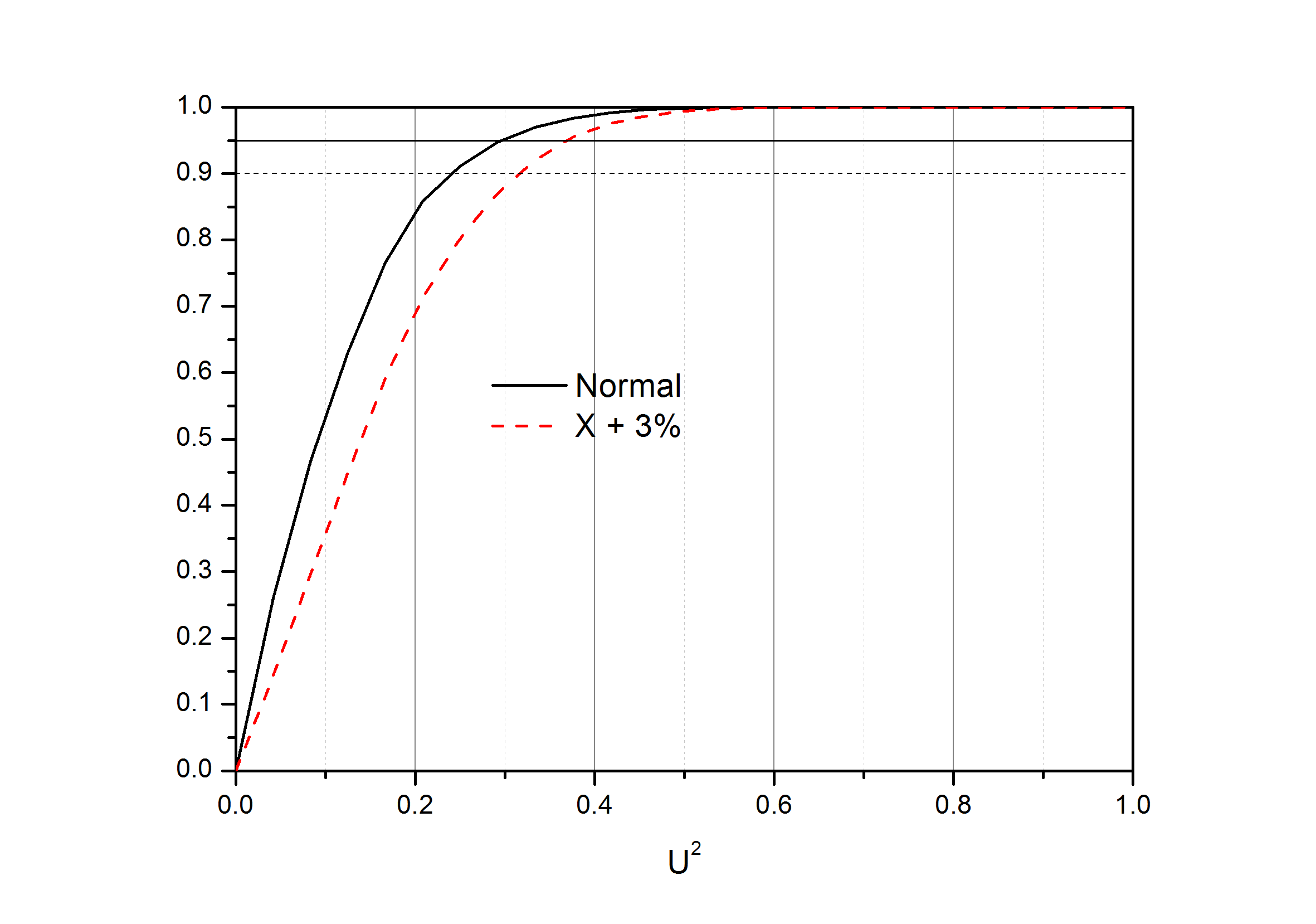}
\end{minipage}
{\caption{The integrated likelihood functions for mass $m_4=2\,eV$ (left) and $m_4=3\,eV$ (right). Solid curve is for the nominal tritium source thickness and dashed curve is for the thickness increased by systematic uncertainty of 3\%. The cross with horizontal lines at 0.9 or 0.95 defines the relevant upper limit for $U^2_{e4}$.  }\label{fig:syst4}}
\end{figure}

\begin{figure}[h]
\center\includegraphics[width = 0.48\linewidth]{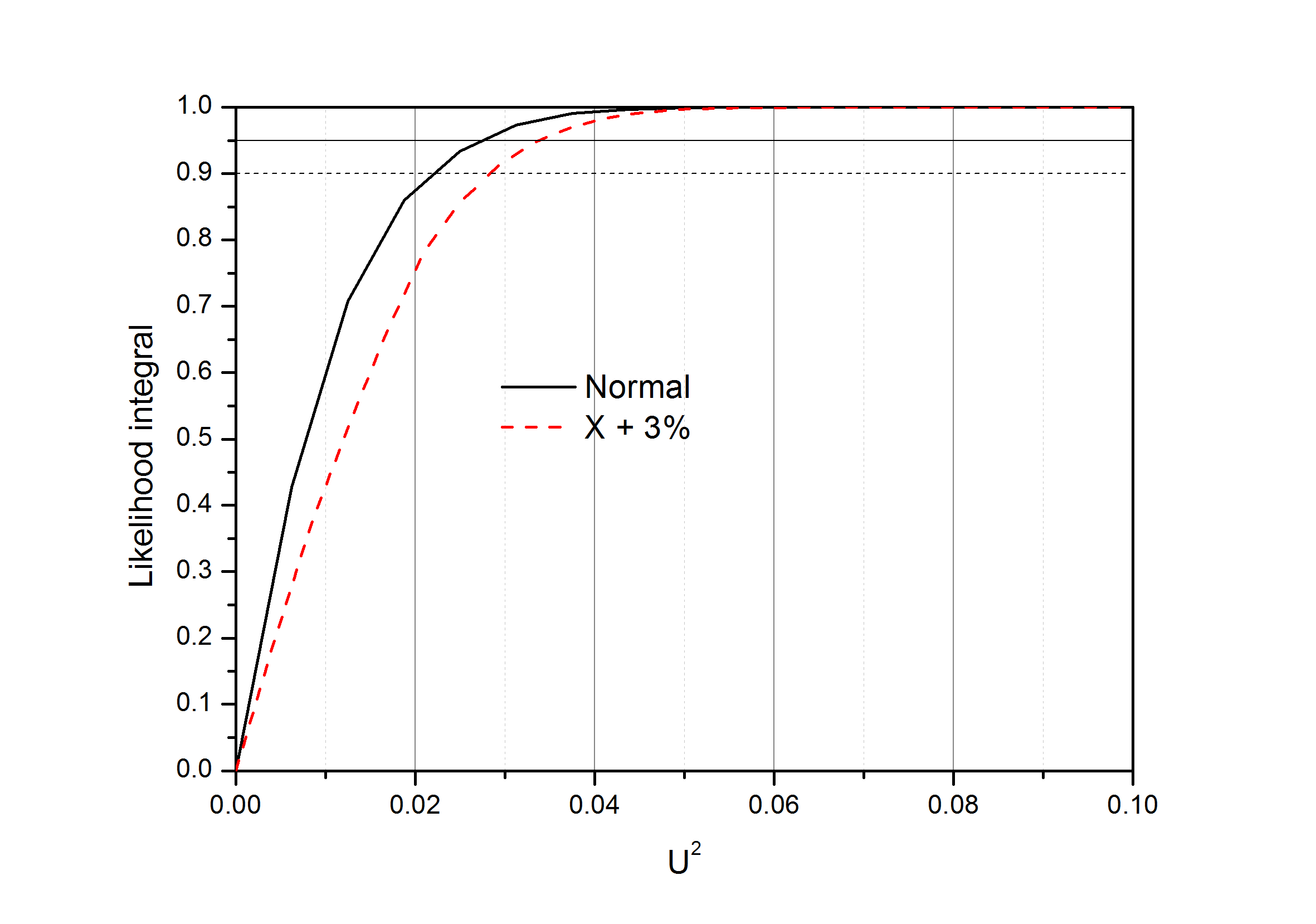}
\caption{Identical to Fig.~\ref{fig:syst4} but for $m_4=10\,eV$}
\label{fig:syst100}
\end{figure}

In the analysis all main systematic uncertainties were checked, which contribute to the systematic error based on our previous publication~\cite{our_nu_e} on the electron neutrino mass, namely:
\begin{enumerate}
\item Uncertainty of the effective gaseous tritium source thickness of 3\%.
\item Usage of different final states spectrum for decay daughter ion of T$^3$He$^+$.
\item Uncertainty of 20\% in the electron trapping effect in the tritium source, see Ref.~\cite{our_nu_e} for details.
\end{enumerate}
The systematic shift due to these effects appeared to be rather small. In Fig.~\ref{fig:syst4} and   Fig.~\ref{fig:syst100}  the difference of the integrated likelihood functions for masses 2\,eV, 3\,eV and 10\,eV are shown at nominal and shifted by a +3\%  source thickness, thus it gives a conservative estimation. Doing a $\pm$3\% variation of the  source thickness we select the largest variation. The average shift of the upper limit in systematics estimation procedure is about 12\%. At 95\% the shift of the upper limit of the confidence level (the curve crossing with the horizontal line at 0.95) is rather small even for a conservative systematic influence estimation.

\begin{figure}[h]
\begin{minipage}[t]{0.48\linewidth}
\includegraphics[width=1.\textwidth]{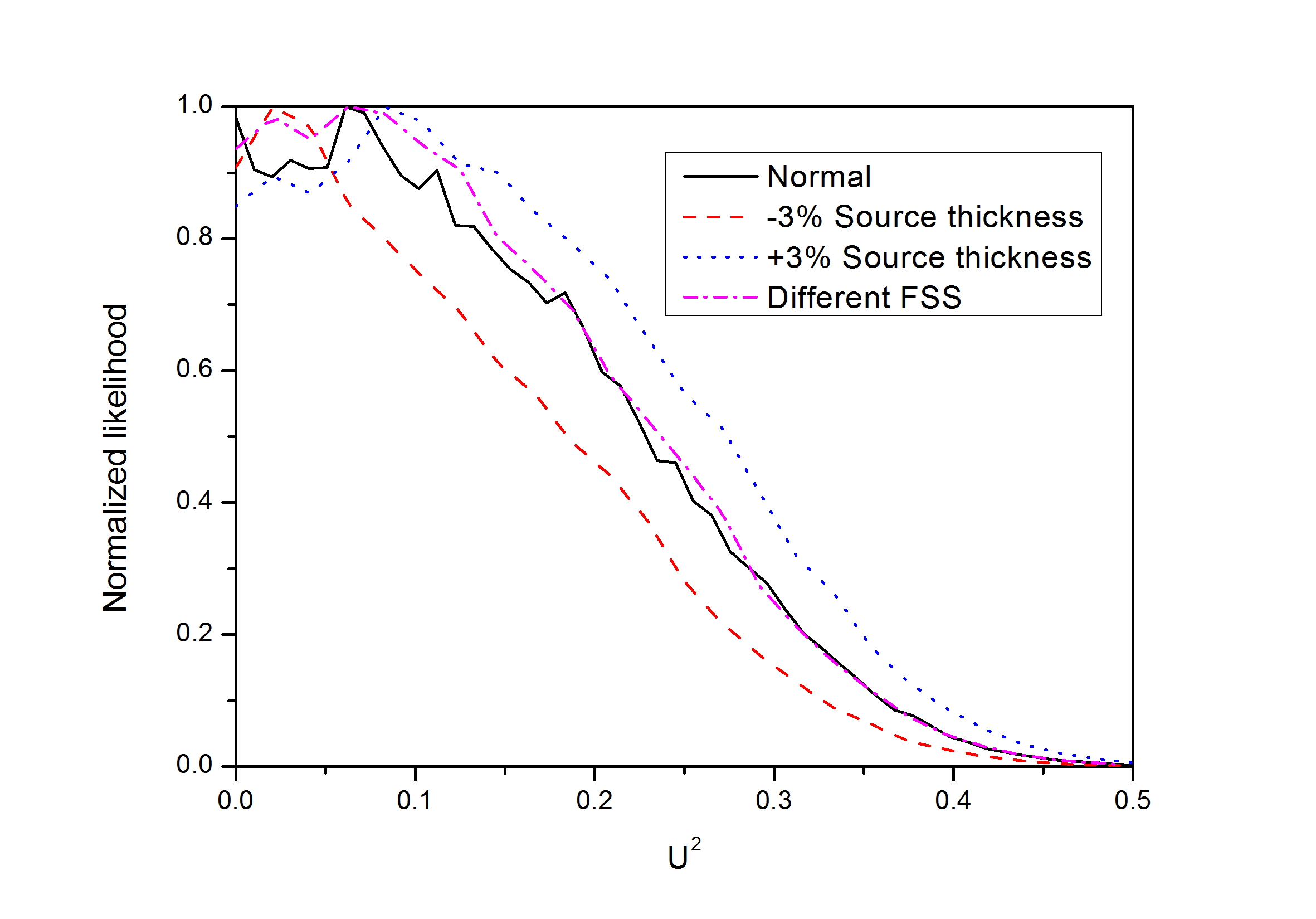}
\end{minipage}
\hfill
\begin{minipage}[t]{0.48\linewidth}
\includegraphics[width=1.\textwidth]{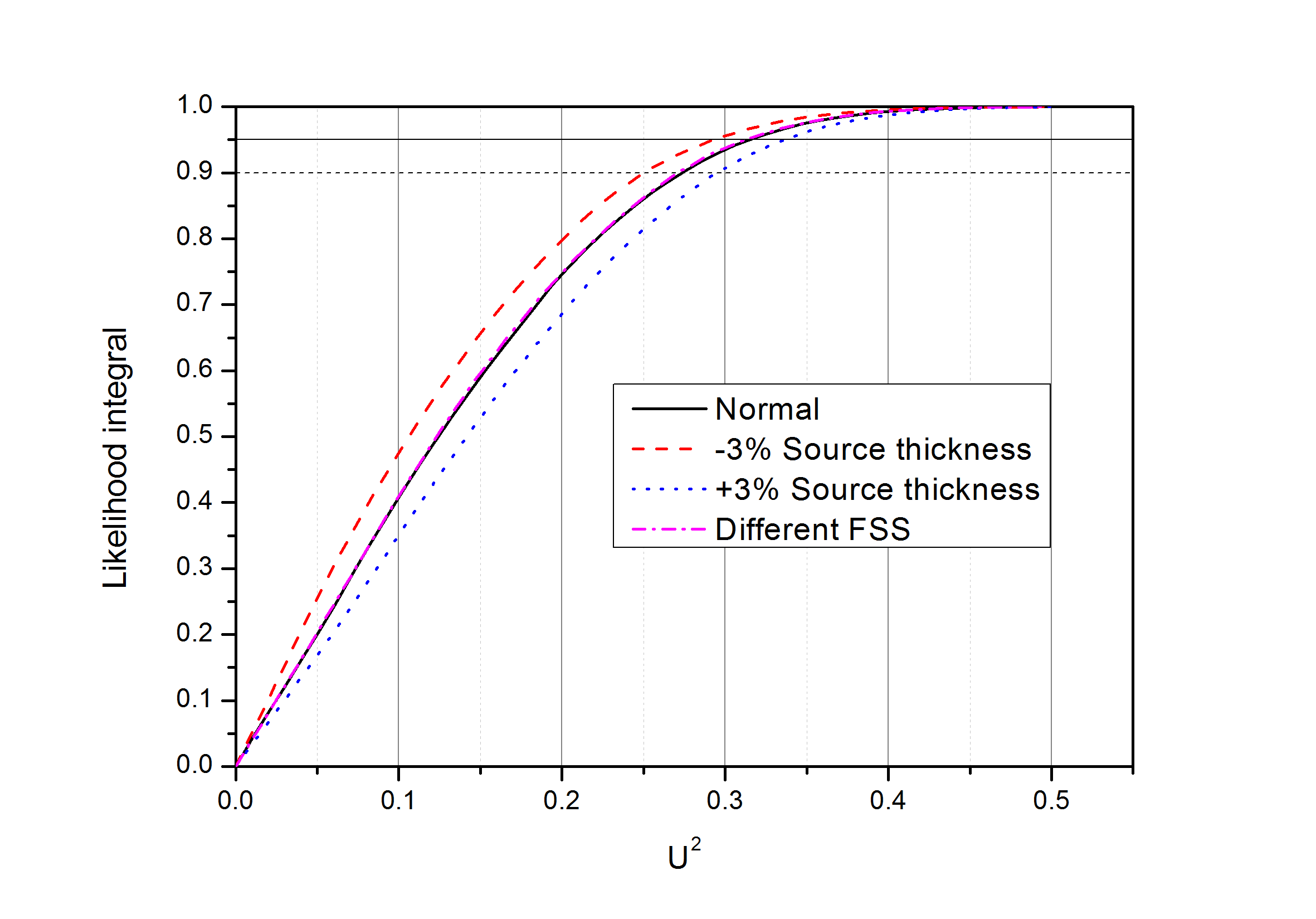}
\end{minipage}
{\caption{Distribution of likelihood function for $m_4=5\,eV$  and Run28 with variation of different systematic parameters (left) and the relevant integrated likelihood functions (right). }
\label{fig:run28_5}}
\end{figure}
\begin{figure}[h]
\begin{minipage}[t]{0.48\linewidth}
\includegraphics[width=1.\textwidth]{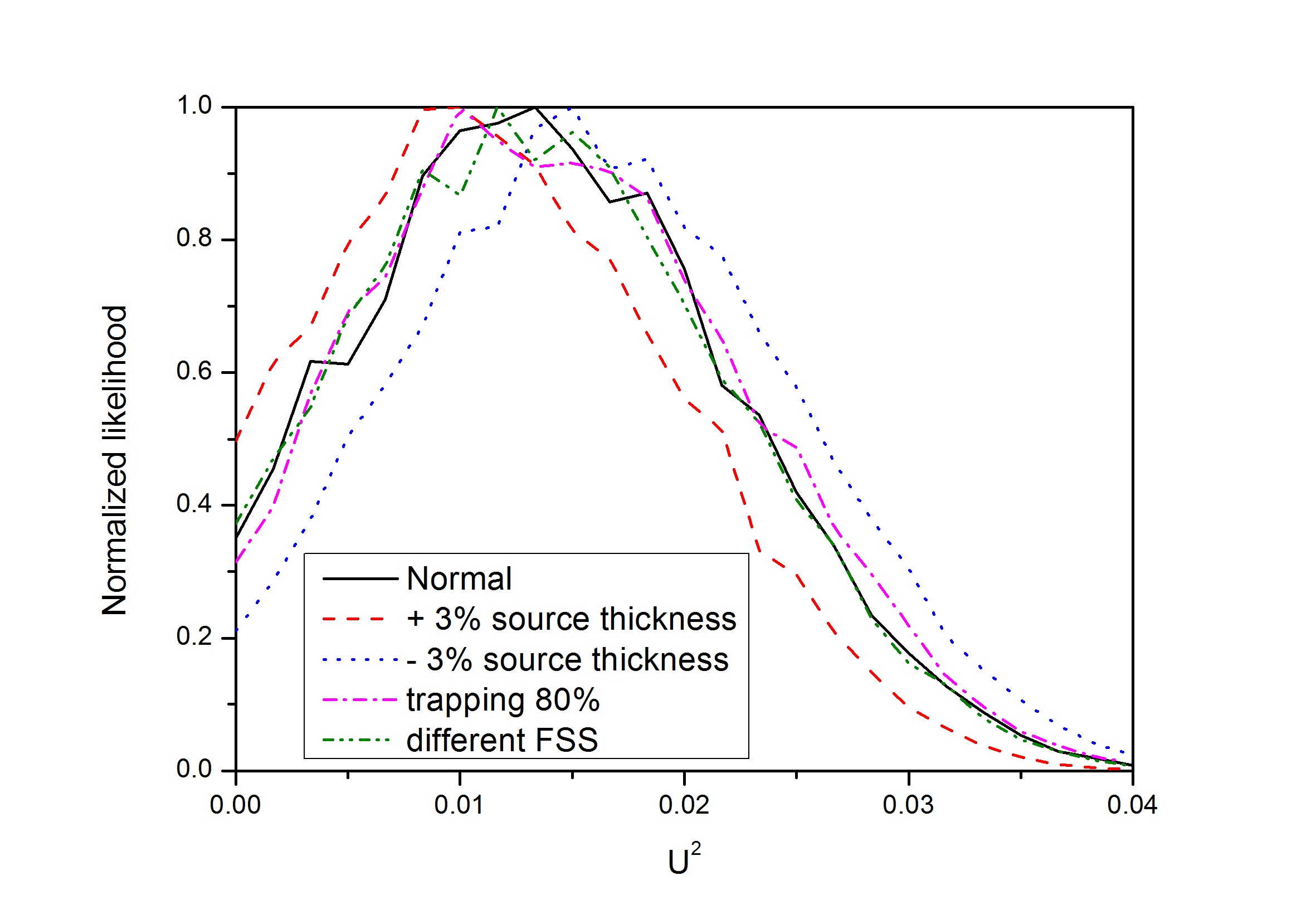}
\end{minipage}
\hfill
\begin{minipage}[t]{0.48\linewidth}
\includegraphics[width=1.\textwidth]{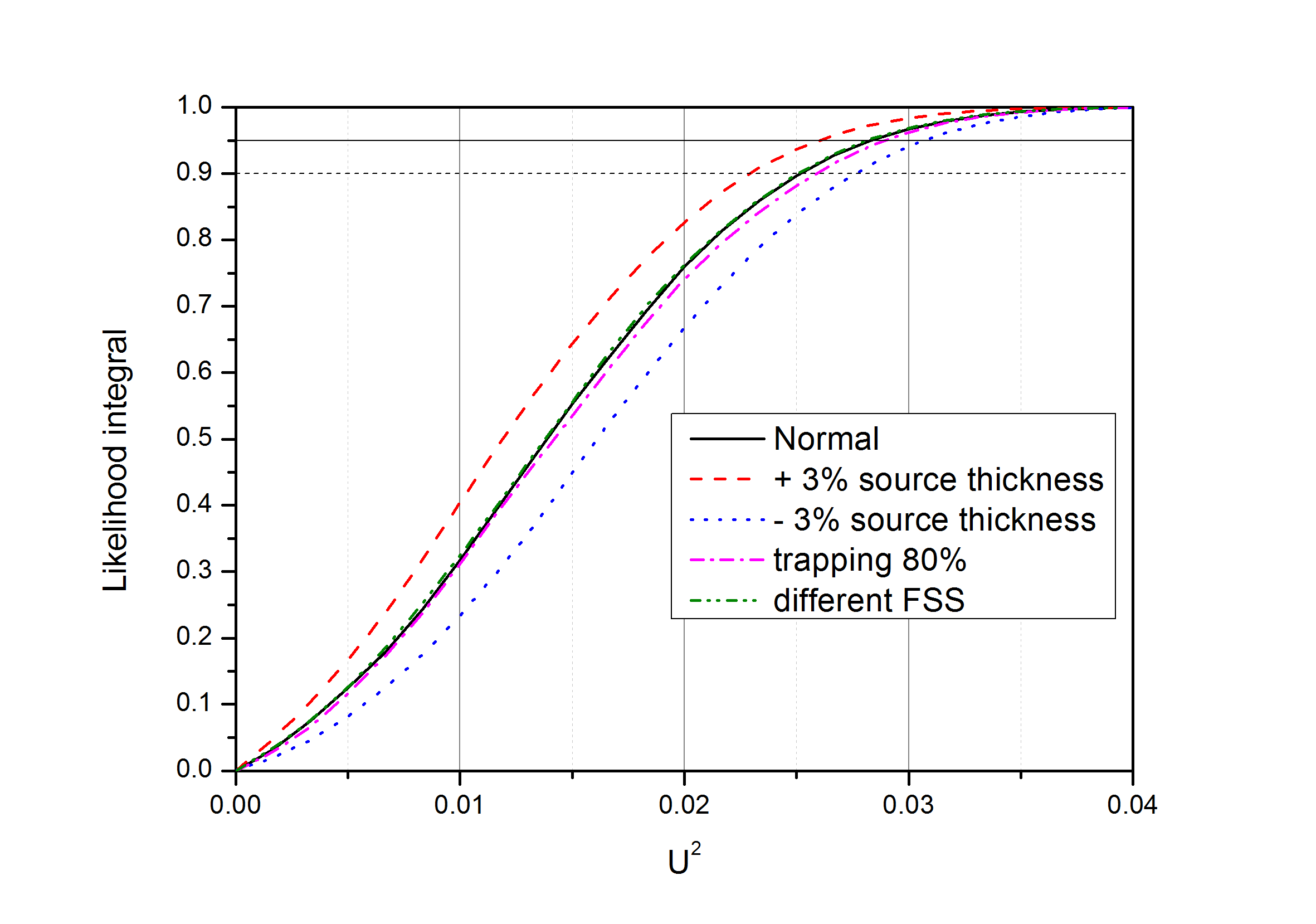}
\end{minipage}
{\caption{Distribution of likelihood function for $m_4=50\,eV$  and Run24 with variation of different systematic parameters (left) and the relevant integrated likelihood functions (right).}
\label{fig:run24_2500}}
\end{figure}
The effect of other systematic parameters is even less. It is clearly seen in Fig.~\ref{fig:run28_5} and Fig.~\ref{fig:run24_2500}, where changes of the likelihood function by varying  systematic parameters are shown for two experimental runs. The reasonable question is why systematic effect from the source thickness has little influece on the upper limit of an additional heavy mass while for the electron antineutrino mass, $m_e$, published in ref.~\cite{our_nu_e} it is comparable with the statistical errors (in ref.~\cite{our_nu_e} for the square of the effective mass we got a statistical error of 1.89 $eV^2$ and a systematical error from the source thickness of the order of 1.68 $eV^2$)? The general answer is that we use two different functions to describe the spectrum shape for the case of electron neutrino mass and for the case when  light mass is assigned to zero  with an additional component for the heavy mass admixture. All systematic errors contribute differently and therefore the effect, say, of the source thickness can be different.
\section{Results and discussion}
\begin{figure}[h]
\begin{minipage}[t]{0.48\linewidth}
\includegraphics[width=1.\textwidth]{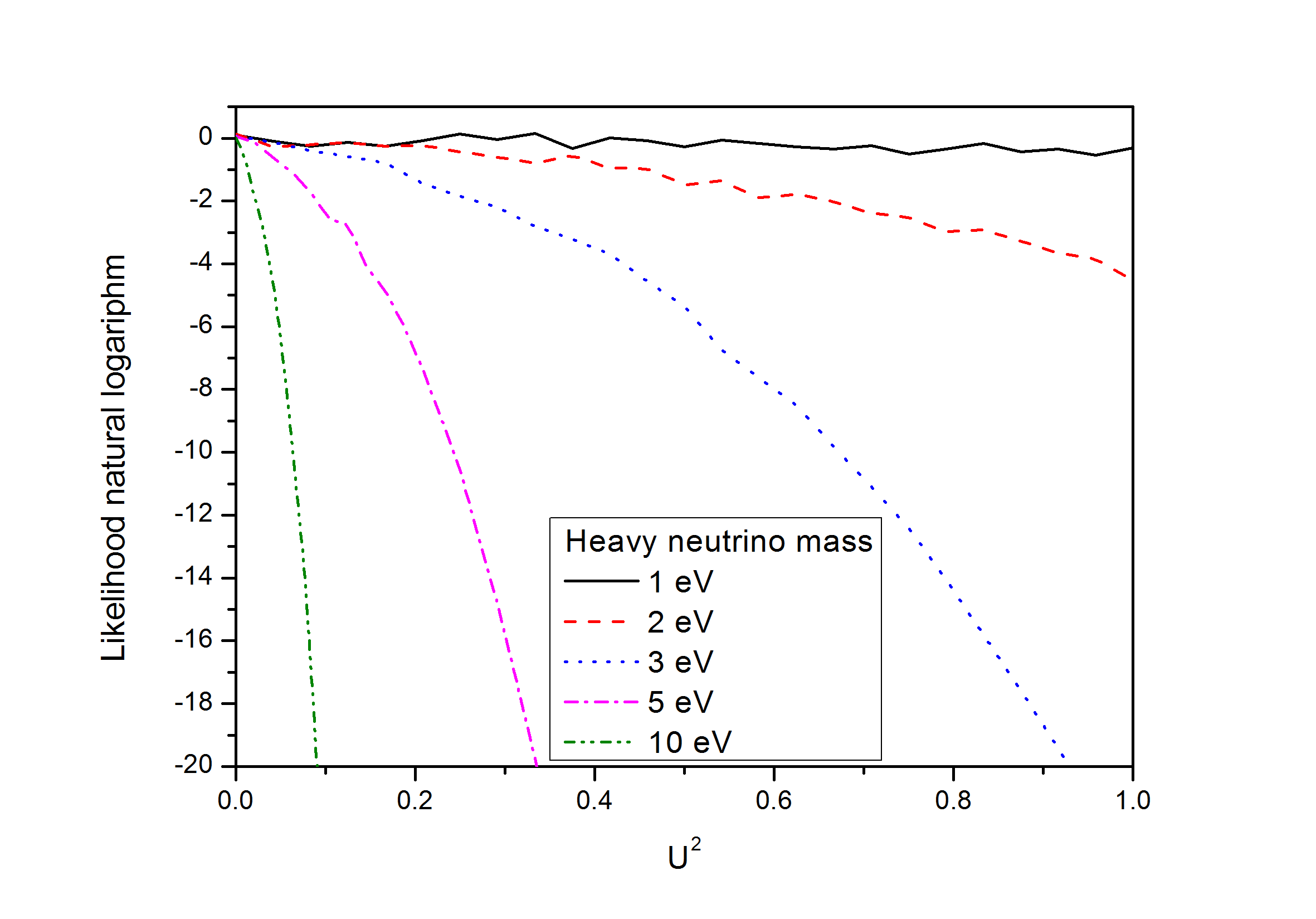}
\end{minipage}
\hfill
\begin{minipage}[t]{0.48\linewidth}
\includegraphics[width=1.\textwidth]{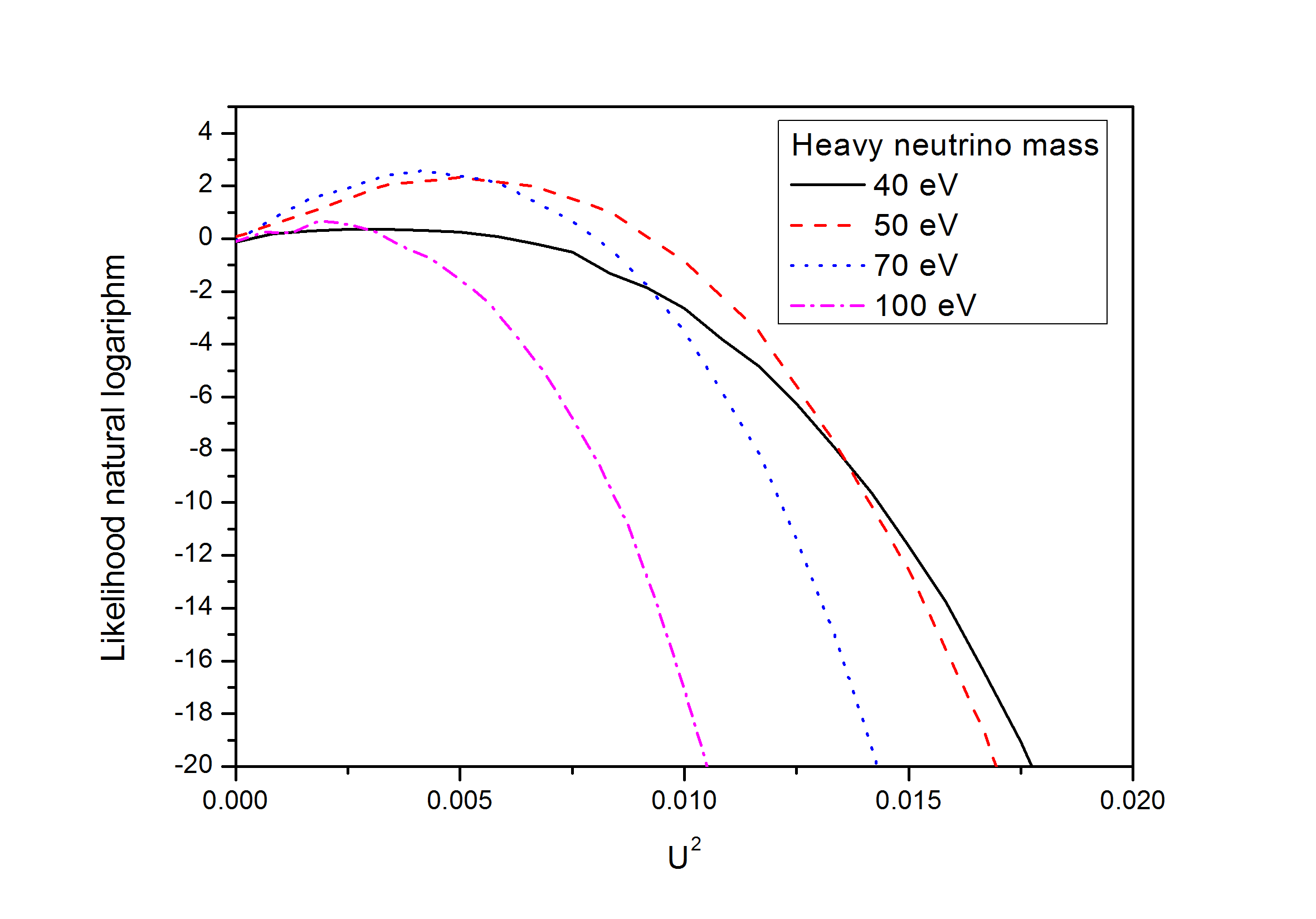}
\end{minipage}
{\caption{The likelihood curves for $U^2$ at different  additional neutrino mass eigenstates, $m_4$. }\label{fig:LnLike1}}
\end{figure}

\begin{figure}[h]
\center{ \includegraphics[width = 100 mm]{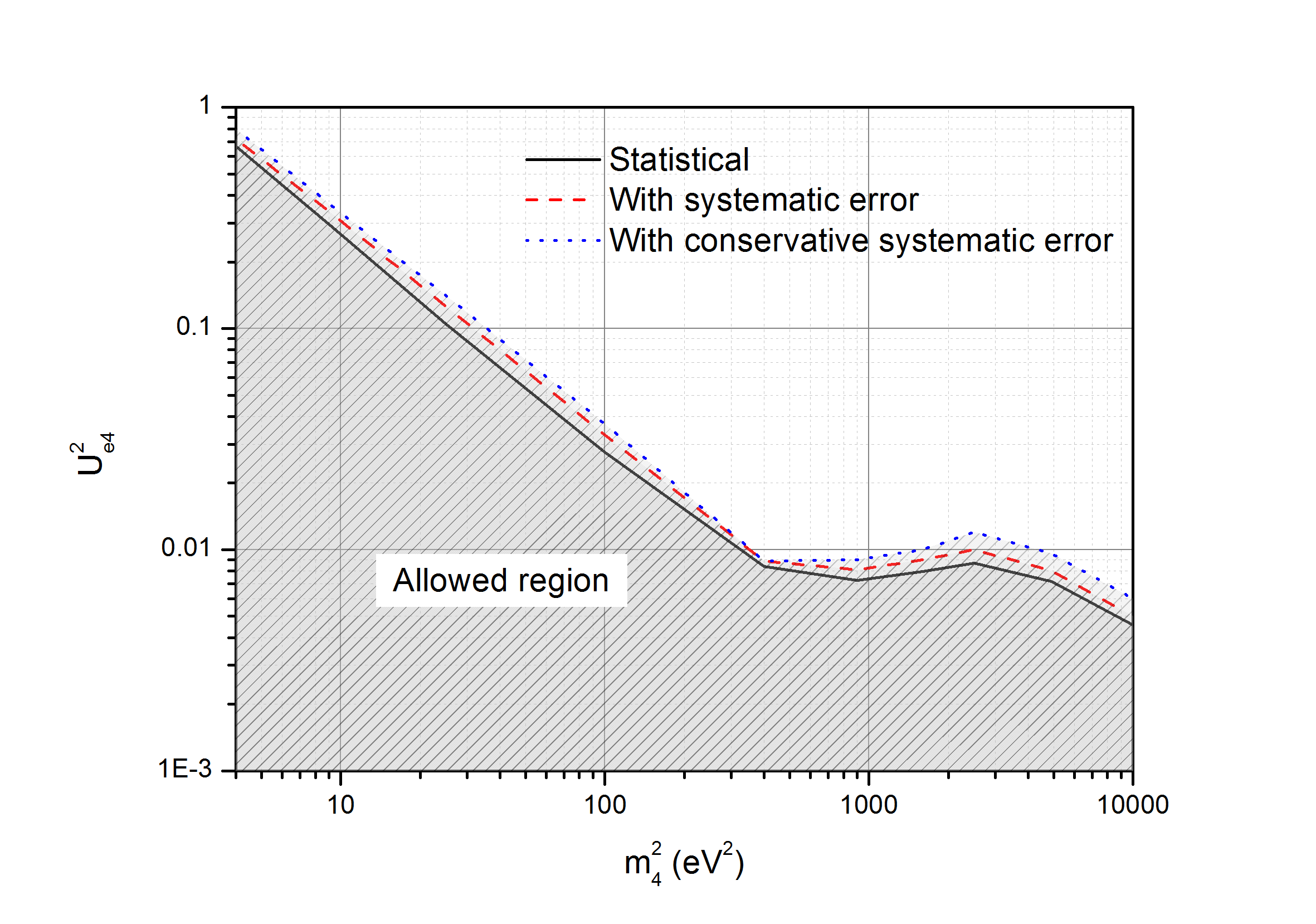}}
\caption{The upper limit at the 95\% C.L. for the square of an additional neutrino mass eigenstate, $m_4^2$. The solid line corresponds to statistical errors only, the dashed line includes estimation with systematic errors, while the dotted line depicts most conservative systematic error estimate.}
\label{fig:limits}
\end{figure}

The ultimate likelihood curves (without systematic influence) for different masses are presented in Fig.~\ref{fig:LnLike1}. The resulting upper limits at a 95\% C.L. for an additional neutrino mass eigenstate, $m_4$, are presented in Fig.~\ref{fig:limits}. The solid line denotes statistical limit, the dashed line corresponds to the normal systematic evaluation and the dotted line corresponds to the conservative systematic estimation. At the mass range $m_4$ from 20\,eV to 100\,eV the upper limit stays between 0.01 and 0.005. As expected, the experiment with a sensitivity limit of about 2\,eV~\cite{our_nu_e} has a poor rejection factor at $m_4$ of about a few electron-volts. 
We exclude point at 1\,eV to avoid misinterpritation with $U^2$=0.95 where $U^2$ saturates for very small $m_4$ as it should be for a flat distribution of the likelihood function, see Fig.~\ref{fig:LnLike1}. We have also to mention that in the first short publication~\cite{short} the  upper limits were defined by statistical errors. That was done to avoid many discussion details of systematic errors which are (as can be seen in Fig.~\ref{fig:limits}) small indeed.  

\subsection{Method cross check for the electron antineutrino case}

\begin{figure}[h]
\center\includegraphics[width = 100 mm]{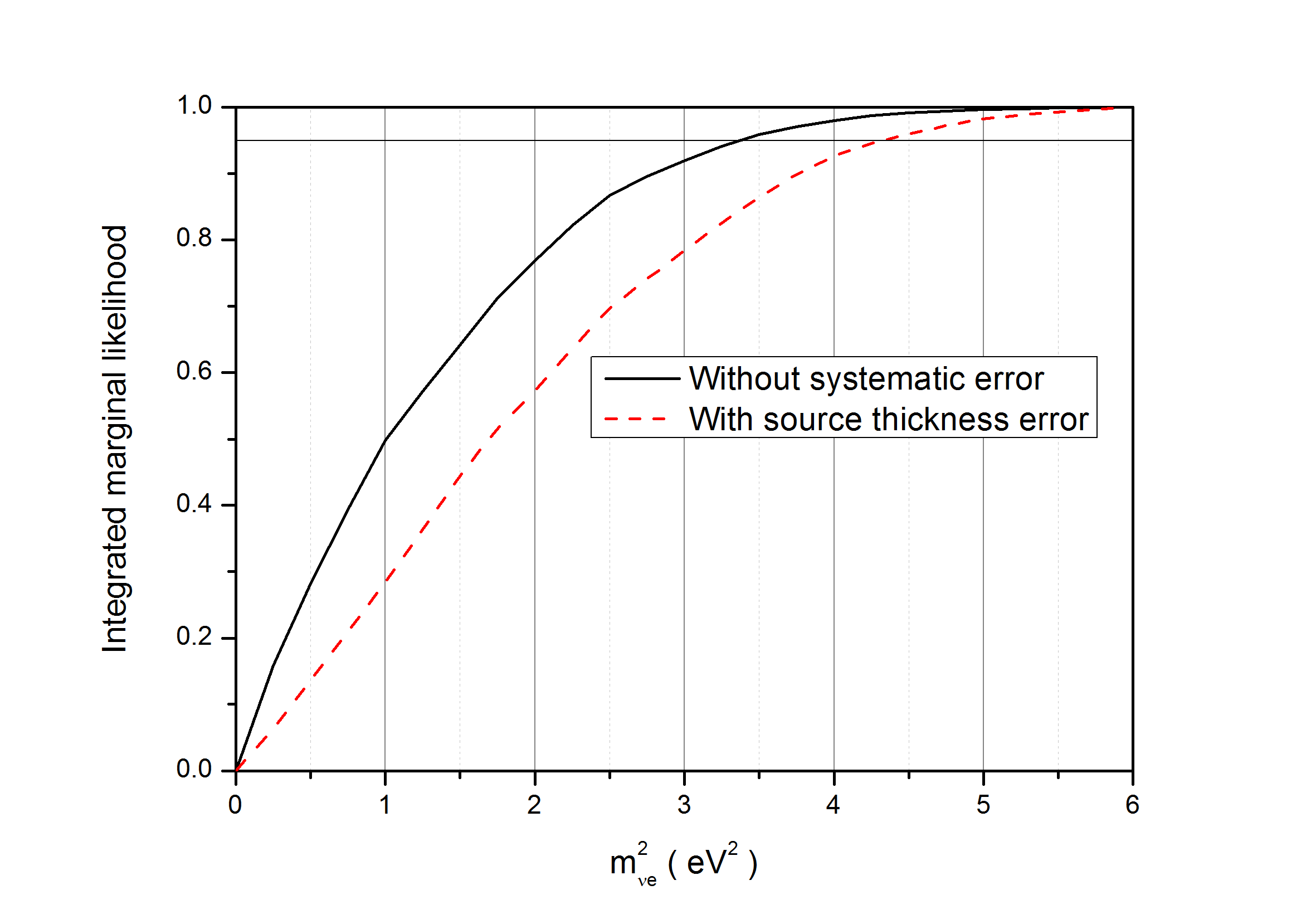}
\caption{The integrated marginal likelihood functions for electron antineutrino squared mass. The solid curve corresponds to the contribution of statistical errors and the dashed like -- to the additional contribution of major systematic uncertainty in the source thickness.}
\label{fig:mnuTest}
\end{figure}

The implemented method proved to be not so much time consuming and much more reliable than the convenient Gaussian techniques. In our paper~\cite{our_nu_e} concerning the results of the electron neutrino mass, $m_{\nu e}$, we applied a commonly used method by explicitly fitting the spectrum. In order to check correspondence to that publication the data were also analyzed with respect to the electron antineutrino mass by using the likelihood function as described in this work. The resulting marginal likelihood integrals for statistical errors only and for  the major systematic contribution are shown in Fig.~\ref{fig:mnuTest} for parameter $m_{\nu e}^2$. It is clearly seen that the upper limit on electron antineutrino mass of $\sqrt{4.35} = 2.08$\,eV is in a good agreement with that presented in ref.~\cite{our_nu_e}, namely 2.12\,eV (for Bayesian method).

\subsection{Problem at 50\,eV heavy mass}

As can be seen in Fig.~\ref{fig:limits}, for $m_4 $ around 50\,eV the upper limit goes up. Moreover, there can be a low bound: in Fig.~\ref{fig:LnLike1} some functions peak at non-zero value of $U^2$.  
At $m_4$=50\,eV the combined likelihood function over all runs peaks at 0.0049 and has a Gaussian width of 0.002. We investigated what could be the origin of such a peak.

The most suspicious element is the final states spectrum of the daughter molecule $T^3He$. As we  have already mentioned, the problem with FSS is that this spectrum has only been calculated and never experimentally checked. We manually added a single line around 50\,eV with an amplitude of 0.5-1\% to FSS (see Fig.2 in \cite{our_nu_e}). Such an addition could be attributed, for example, to the double ionization of $^3$He molecule at the energy of 13.6$\cdot$4=54.4\,eV.  The peak of the likelihood function at $m_4 $ =50\,eV disappears. Thus, the problem could be in  FSS.

Another possibility for appearance of the effect may come at the stage of our raw data processing. An important feature of the procedure is the so called a bunch rejection algorithm. The purpose of such a rejection is to filter short-timed high intensity "bunch" events. They are caused by the electrons trapped in the spectrometer magnetic bottle. The procedure for a "bunch" search is automatic and its effectiveness is lower for higher count rates. The count rate in the spectrometer below the spectrum endpoint by about 50\,eV is usually critical for bunch rejection algorithm. Beyond that point we use a simple extrapolation of bunch contribution. Thus, some non-perfection of the bunch rejection algorithm may distort the spectrum.

\subsection{Comparison with other methods}

In this work we used the marginal likelihood Bayesian procedure to set an upper limit on heavy neutrino mass mixing. The other well known approach to work with parameters near their physical border is the so called frequentist confidence interval estimation. It requires to extend the fit function to unphysical region and then to use the unified approach of Feldman and Cousins~\cite{FeldmanCousins} or a similar technique to place an upper limit on the parameter being studied. The direct fit approach is somewhat easier to use in most cases  because it requires to calculate likelihood or $\chi ^2$ functions only few times instead of covering the whole many-dimensional region of parameters. The weakness of such a method is that one needs to add an additional information by providing fit function extension to the unphysical region. Another flaw is that most of frequentist methods provide correct estimations for errors only in the case of Gaussian likelihood shape. As it was demonstrated in Fig.~\ref{fig:like2d}, sometimes distributions are well beyond any reasonable shape. The Bayesian method is preferred because of its transparency from a mathematical point of view and its additional flexibility (for example it allows one to work with non-Gaussian distributions).

\begin{figure}[ht]
\center\includegraphics[width = 100 mm]{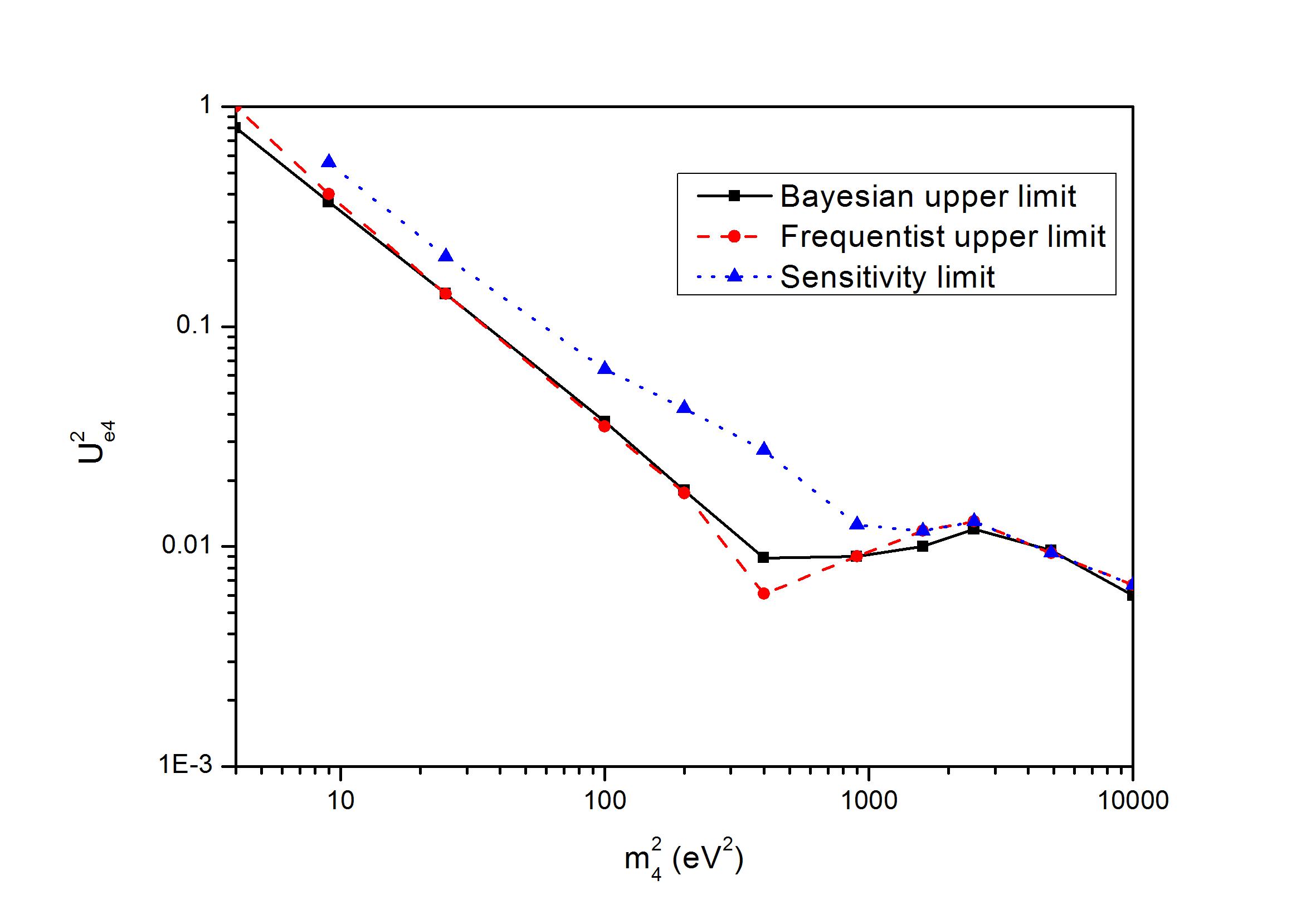}
\caption{Comparison of 95\% C. L. upper limits with different methods: Bayesian approach used in the analysis (the most conservative one), frequentist or Feldman and Cousins and the range defined by a sensitivity limit.}
\label{fig:compare-methods}
\end{figure}

We explicitly checked that the difference in the upper limits obtained by our procedure and standard Gaussian fit (JMINUIT~\cite{jminuit} and quasi-optimal weights~\cite{QOW} were used) does not exceed 10-20\%. In Fig.~\ref{fig:compare-methods}  Bayesian upper limits are plotted next to similar estimates by  Feldman and Cousins procedure. Point by point results for standard Gaussian fit are plotted in Fig.~\ref{fig:troitsk-mainz} by solid dots. In Fig.~\ref{fig:compare-methods} we also presented the sensitivity limit calculated in the same way as it was estimated in Ref.~\cite{our_nu_e}: in case when the fit gives a negative value, the sensitivity limit is  simply estimated by setting that value at zero and keeping the same error bars. 

\begin{figure}[htb]
\center\includegraphics[width = 100 mm]{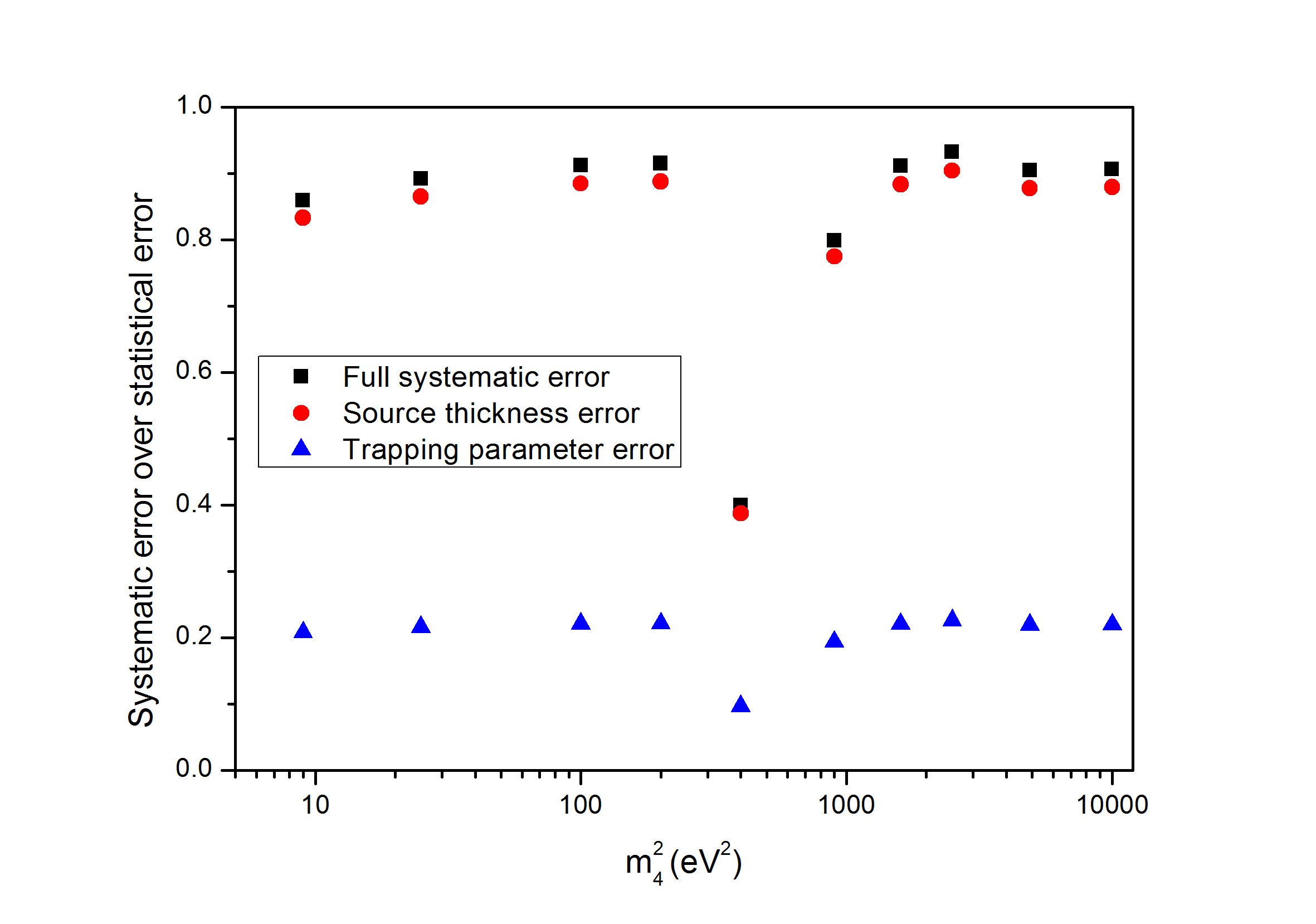}
\caption{The systematic error of $U^2$ in Feldman and Cousins approach from different sources divided by statistical error for different masses. }
\label{fig:systematic}
\end{figure}

The influence of systematic errors in calculating Feldman and Cousins upper limit is shown in Fig.~\ref{fig:systematic}. In this case the total span of systematic parameters ($\pm$ 3\% for source thickness and $\pm$ 20\% for trapping) was taken as a standard deviation of its distribution. The drop of systematic error for a squared mass 400 is caused by the fact that the influence of systematic parameters changes its sign in this region (for small masses an increase at source thickness causes an increase in $U^2$, but for higher masses the effect is reverse). As one can see, the systematic uncertainly is in the range of 0.8--0.9 of the statistical error. Thus, the total uncertainty increases by factor 1.3--1.35.

One cannot directly compare systematic influence in marginal likelihood Bayesian method and Feldman and Cousins approach because the estimation of upper limit itself is obtained by other means. There is no such  thing as parameter error in marginal likelihood method because there is no point parameter estimation. One does not use the central value and its error and instead takes the whole likelihood shape as it is done in Fig.~\ref{fig:syst4} and Fig.~\ref{fig:syst100}.

\subsection{Error estimation from the limit for electron antineutrino mass}

It is worthwhile to estimate the direct fit error bars based on our previous measurements for electron antineutrino mass~\cite{our_nu_e}. 
In both analyses -- for the active neutrino case and for the search for an additional eigenstate -- the same electron spectrum is fitted by two different functions with different shapes. This provides different results, but some connection for small values of $m_4$ could be established.
Investigating Eq.~\ref{eq:one} one can derive the following spectrum dependence on neutrino mass:
\begin{equation}
S(E) \sim \sqrt{1- \frac{m_\nu^2}{(E_0-E)^2}} \approx 1 - \frac{m_\nu^2}{2 (E_0-E)^2}.
\end{equation}
The expansion is valid only for small masses but could be applied in the case of neutrino mixing (Eq.~\ref{eq:funct}):
\begin{equation}
S(E) \sim (1-U_{e4}^2) + U_{e4}^2  \sqrt{( 1 - \frac{m_4^2}{(E_0-E)^2}} \approx 1 - \frac{U_{e4}^2 m_4^2}{2 (E_0-E)^2}.
\end{equation}
Comparing these equations one can derive:
\begin{equation}
m_{\nu} \approx U_{e4}^2 m_4^2 
\label{eq:light_heavy}
\end{equation}
Thus, we get a simple relation between the errors:
\begin{equation}
\sigma (U_{e4}^2) \approx \frac{\sigma( m_{\nu}^2)}{m_4^2}.
\label{eq:light_heavy_error}
\end{equation}

\begin{figure}[h]
\center\includegraphics[width = 100 mm]{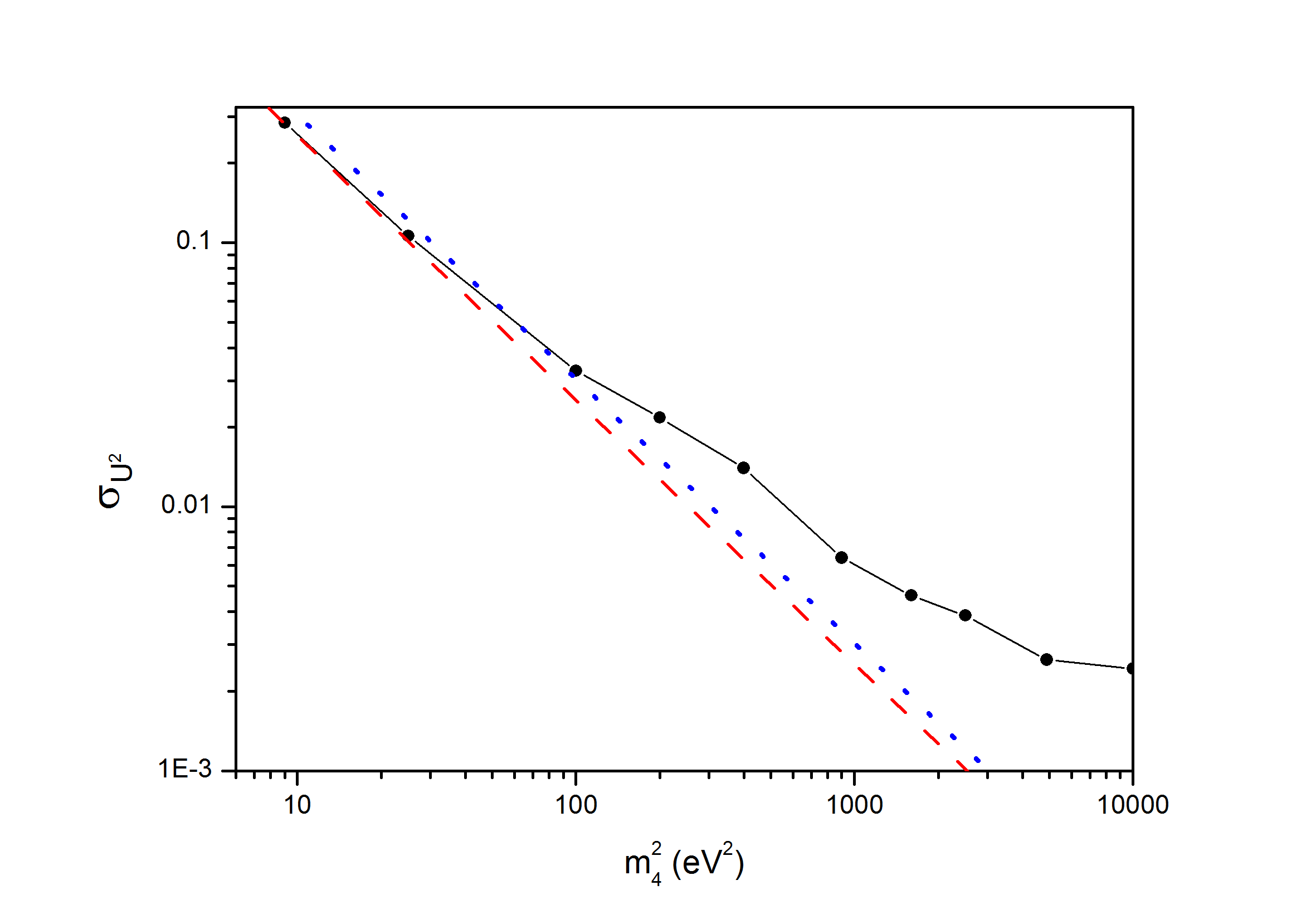}
\caption{(color online) Comparison of errors for heavy neutrino mass obtained by the current analysis, black symbols connected by solid lines, and approximate estimation $\sigma (U_{e4}^2) = 2.53/m_4^2$ based on the rusult for the electron antineutrino mass~\cite{our_nu_e}, red dashed line. The blue dotted line corresponds to the estimation $\sigma (U_{e4}^2) = 3.04/m_4^2$ for the total error from Ref.~\cite{mainz2}.} 
\label{fig:light_heavy}
\end{figure}

While these relations are rather approximate they could be used to compare the results for the cases with antineutrino mass and heavy neutrino at a small value of mass. 
In Ref.~\cite{our_nu_e} in the analysis for active neutrino the total error was $\sigma ( m_{\nu}^2)=2.53\, eV^2$. 
In Fig.~\ref{fig:light_heavy} the red dashed straight line corresponds to the equation $\sigma (U_{e4}^2) = 2.53/m_4^2$ and represents the expectation based on the previous result. The black dots connected by the solid lines correspond to the errors obtained by the current analysis, see error bars in Fig.~\ref{fig:troitsk-mainz}. One can see a very good agreement at small $m_4^2$ . 

\subsection{Comparison with the Mainz result}

Recently a similar analysis has been published by a few members from the Mainz Neutrino group~\cite{mainz-sterile}, where they analyzed  their old data for electron antineutrino mass measurements with a condensed tritium source~\cite{mainz2}. The method which has been used to sum up data from all runs is very similar to that we used. The difference is that they were summing up the $\chi^2$  instead of multiplying the likelihood which should provide the same result in most cases. We described the difference between Bayesian and Feldman and Cousins frequentist approach earlier.

The upper limits reported in~\cite{mainz-sterile} are about two times higher than ours presented in Ref.~\cite{short}. The reasonable question is why our heavy neutrino results are so different while the estimations  of light electron neutrinos are similar. The first reason  is that we have set of dissimilar systematic uncertainties. The systematic error of our experiment comes mostly from the uncertainty of the effective thickness of our gaseous source. As we have already shown the change of the spectrum shape due to the change of the source thickness does not give a large effect on heavy mass mixing estimations. We can't speculate about the systematics of the Mainz experiment but from the resulting figure and table 1 in \cite{mainz-sterile} their errors seem to come mostly from some uncertainties in the quench-condensed source. The spectrum distortion from these sources is definitely different compared to that from a change of gaseous source thickness. 

\begin{figure}[h]
\center\includegraphics[width = 100 mm]{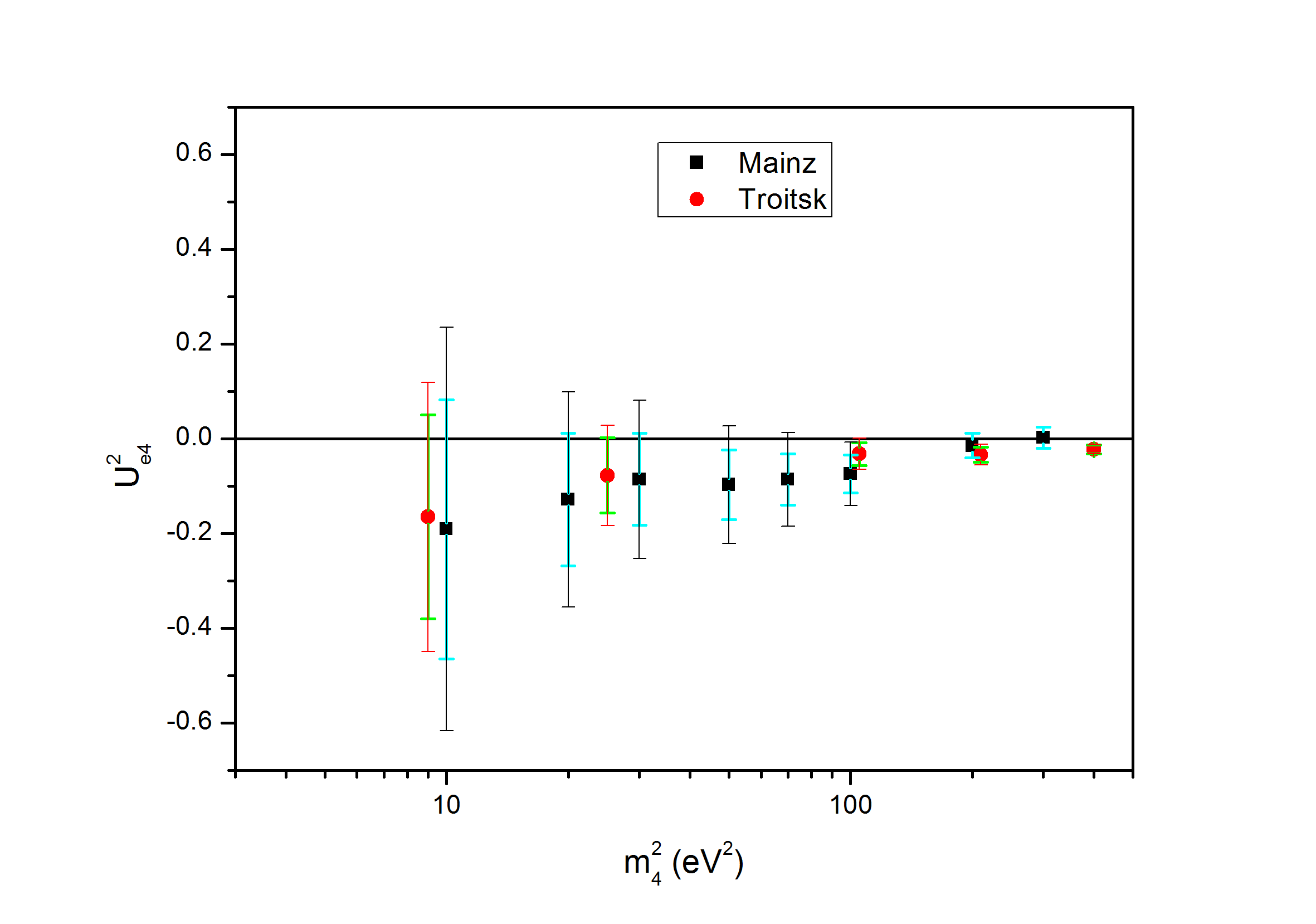}
\caption{Comparison of our and Mainz frequentist results. The inner error bars correspond to statistical errors, the outer ones -- to total errors. Points for squared masses 100 and 200 are intentionally slightly shifted  to provide plot readability.} 
\label{fig:troitsk-mainz}
\end{figure}

\begin{figure}[h]
\center\includegraphics[width = 100 mm]{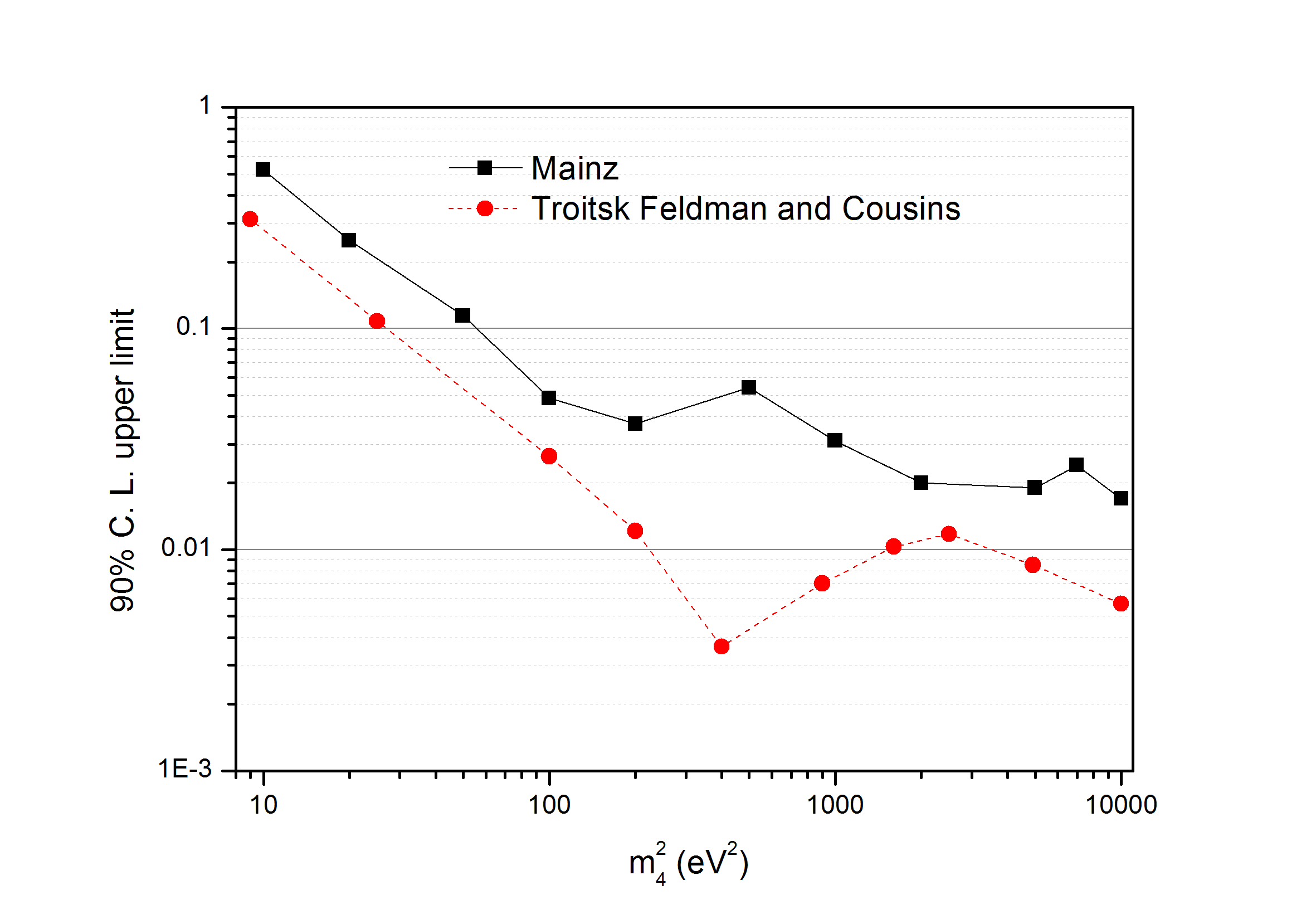}
\caption{Comparison of Troitsk and Mainz 90\% C. L. upper limits obtained by Feldman and Cousins approach.} 
\label{fig:troitsk-mainz-limits}
\end{figure}

The second reason is some difference in statistical errors. Again, we can not make any assumption on the amount of data in Mainz analysis but we have also performed frequentist analysis  and present the results in terms of Gaussian average values and errors. We reanalyzed Troitsk data in the same manner as the Mainz Neutrino group~\cite{mainz-sterile} did, Fig.~\ref{fig:troitsk-mainz}. Comparing our statistical and systematic errors with those taken from Figure 4 in~\cite{mainz-sterile}, we see that in most cases we have  better a statistical precision and lower systematic influence. To make an exact comparison, in   Fig.~\ref{fig:troitsk-mainz-limits} the 90\% C. L. upper limits obtained by Feldman and Cousins approach are plotted for both experiments using points from Fig.~\ref{fig:troitsk-mainz}. It confirms that the difference is still there, regardless of the analysis method which has been used. 

One can also apply the approach from section D to estimate error bars based on the previous Mainz group measurements for the electron antineutrino mass~\cite{mainz2}. The total error from that reference for $m_{\nu}^2$ is $\sqrt{2.2^2+2.1^2}=3.04\, eV^2$. The corresponding estimate $\sigma (U_{e4}^2) = 3.04/m_4^2$ is plotted in Fig.~\ref{fig:light_heavy} by a blue  dotted line. 
The error estimates for $\sigma(U_{e4}^2)$ from the Troitsk and Mainz experiments are close.  

To top it all, the Mainz group used only the last 70\,eV of the beta spectrum for masses $m^2\,\leq$\,1000\,eV$^2$,  while we always used a wider range of the last 175 eV ($E_{low}$ = 18400\,V).

\section{Conclusion}

In conclusion, we present all details of the analysis of a search for an additional neutrino mass eigenstate. The same data set and the analysis framework of the direct electron antineutrino mass measurements in tritium $\beta$-decay in Troitsk were used.   The maximum likelihood method was applied to evaluate a possible contribution  of the heavy extra mass state  $m_4$ and its amplitude $U^2_{e4}$. Smallness in difference for three active neutrino masses allows one to assume that all three active neutrinos have zero masses. We constructed a four-dimensional likelihood function $L(U^2_{e4}, E_0, N, bkg)$, which evaluates the $\beta$-electron spectrum. Then, likelihood function was marginalized over all non-essential parameters by integrating  them over 3-dimensions: $L(U^2_{e4}) = \int\limits_{E_0}\int\limits_{N}\int\limits_{bkg} L(U^2_{e4}, E_0, N, bkg)$.  The integration is made by the Monte-Carlo procedure. The final posterior probability $L$ for parameter $U^2_{e4}$ is calculated as a product of probabilities $L_k$ for different experimental runs. To estimate a 95\% C.L. for parameter $U^2_{e4}$, $L$ was then integrated over a $U^2_{e4}$ to find  the range containing 0.95 of the total area under  $L$. Systematic errors were estimated for three major sources of the experimental uncertainties: effective thickness of the gaseous tritium source, usage of different  final states spectrum for decay daughter molecule of T$^3$He, uncertainty of 20\% in the electron trapping effect in the tritium source parameter. Interestingly, contributions from these systematic errors are small and do not exceed 15-20\% (for a normal systematic estimation procedure). Finally, all likelihood functions for $U^2$ at different values of additional mass eigenstate $m_4$ were presented and relevant upper limits at a 95\% C.L. were calculated. At the range of $m_4$ from 20\,eV to 100\,eV the upper limit stays between 0.01 and 0.005.  As expected, for masses 2--3\,eV the sensitivity is poor, at larger masses the limit goes down. To check consistency of the maximum likelihood method with the previously used analysis method, we evaluated an upper limit for electron antineutrino mass with the current method and found a good agreement with the previously published value. Current errors for $U^2_{e4}$ were also estimated based on our previous measurements for electron antineutrino mass. At small $m_4^2$ good agreement was observed. 
In the end, a comparison with recent Mainz result has been done. There can be at least two reasons why our result is about two times better. In our case, under assumption that active neutrino mass is zero, systematic uncertainties have smaller effect on an additional heavy mass. At the same time, in analysis for Mainz data systematic uncertainties almost double the total error bars. The comparison  of statistical errors also gives some favor to our result.

The current analysis  was supported by RFBR under grant numbers 11-02-00935-a, 12-02-31323-mol-a and 12-02-12140-ofi-m. We would like to thank  our colleagues I.\,I.~Tkachev and  N.\,A.~Titov for useful discussions.


\section*{References}
\begin{thebibliography}{10}
\bibitem{short}A.\,I. Belesev {\it et. al.}, JETP Lett. {\bf 97}, 67 (2013) [arXiv:1211.7193 [hep-ex]].
\bibitem{our_nu_e}V.\,N. Aseev {\it et. al.}, Phys. Rev. {\bf D84}, 112003 (2011) [arXiv:1108.5034 [hep-ex]].
\bibitem{mainz-sterile} C.~Kraus, A.~Singer, K.~Valerius and C.~Weinheimer,  Eur.\ Phys.\ J.\ C {\bf 73}, 2323 (2013) [arXiv:1210.4194 [hep-ex]].
\bibitem{oscill} K.~Nakamura et al., [Particle Data Group], J.~Phys. G {\bf 37}, 075021 (2010).
\bibitem{FeldmanCousins} G.~J.~Feldman and R.~D.~Cousins,  Phys.\ Rev.\ D {\bf 57}, 3873 (1998) [physics/9711021 [physics.data-an]].
\bibitem{jminuit} http://java.freehep.org/freehep-jminuit/
\bibitem{QOW}F.\,V. Tkachov, arXiv:physics/0604127 (2007).
\bibitem{mainz2} C.~Kraus, {\it et al.}, Eur.\ Phys.\ J.\ C {\bf 40}, 447 (2005) [hep-ex/0412056 (2004)].
\end{thebibliography}
\end{document}